# Omnidirectional 3D printing of PEDOT:PSS aerogels with tunable electromechanical performance for unconventional stretchable interconnects and thermoelectrics

Hasan Emre Baysal,[a] Tzu-Yi Yu,[a] Viktor Naenen,[a] Stijn De Smedt,[b] Defne Hiz,[a] Bokai Zhang,[a] Heyi Xia,[a] Isidro Florenciano,[a] Martin Rosenthal,[c] Ruth Cardinaels,[b] and Francisco Molina-Lopez*[a]

The next generation of soft electronics will expand to the third dimension. This will require the integration of mechanically-compliant three-dimensional functional structures with stretchable materials. This study demonstrates omnidirectional direct ink writing (DIW) of Poly(3,4-ethylenedioxythiophene) polystyrene sulfonate (PEDOT:PSS) aerogels with tunable electrical and mechanical performance, which can be integrated with soft substrates. Several PEDOT:PSS hydrogels were formulated for DIW and freeze-dried directly on stretchable substrates to form integrated aerogels displaying high shape fidelity and minimal shrinkage. The effect of additives and processing in the PEDOT:PSS hydro and aerogels morphology, and the link with their electromechanical properties was elucidated. This technology demonstrated 3D-structured stretchable interconnects and planar thermoelectric generators (TEGs) for skin electronics, as well as vertically-printed high aspect ratio thermoelectric pillars with a high ZT value of $3.2 \cdot 10^{-3}$ and ultra-low thermal conductivity of 0.065 W/(m K). Despite their comparatively low ZT, the aerogel pillars outpowered their dense counterparts in realistic energy harvesting scenarios where contact resistances cannot be ignored, and produced up to 26 nW/cm$^2$ (corresponding to a gravimetric power density of 0.76 mW/kg) for a difference of temperature of 15 K. This work suggests promising advancements in soft and energy-efficiency electronic systems relevant to soft robotics and wearable.

## Introduction

Standing at the intersection of electrical engineering and materials science, soft electronics refers to a burgeoning class of electronic systems that share the form factor of biological systems[1], i. e. they are mechanically soft and stretchable, and can take up complex shapes. Soft electronics offer tremendous potential for application in soft robotics[2], wearable electronics[3–6], biomedical devices[4,7] and tissue engineering[8]. Unlike traditional electronics, soft circuits are not exclusively planar. Hence, they require fabrication techniques that go beyond thin-film fabrication and move toward the third dimension[9–13], a requirement that 3D printing fulfills. In particular direct ink writing (DIW) allows additive manufacturing in three dimensions[10]. In DIW, ink with specific rheology, i.e. shear thinning and high yield stress[10,14], is extruded through a nozzle on a substrate, allowing for the 3D construction of intricate self-supported structures. A robotic deposition system controls the nozzle movement and the material extrusion. The deposited ink must maintain its shape and adhere to the underlying layers to create a cohesive structure. DIW of soft electronic materials such as conjugated polymers[10,15] and hybrid organic-inorganic composites[12] has been reported for applications such as tissue engineering[8], soft robotics[10], electromagnetic shielding[16], and 3D electrical interconnects[13,17].

The steep development of the soft electronics field comes with enhanced functionalities and an ever-increasing energy demand that batteries cannot always fulfill either because they are too bulky or rigid for certain applications or because regular replacement might be inconvenient, unfeasible or environmentally unsustainable. Hence, powering soft electronics requires versatile renewable power sources with compatible unconventional form factors to work along -or instead of- batteries. Thermoelectric (TE) materials can generate electrical power from heat flows. Compared to competing harvesting technologies such as photovoltaics, piezoelectrics, or triboelectrics[6], TEs do not require light or mechanical vibrations or friction to operate, making them ideal candidates to power wearables by exploiting the temperature gradient between the human body and the surrounding environment. The performance of a TE material is evaluated by the dimensionless figure of merit $ZT$, defined as $ZT = \sigma S^2 T/\kappa$, where $\sigma$, $S$, $T$, and $\kappa$ denote electrical conductivity, Seebeck coefficient, absolute temperature, and thermal conductivity, respectively. The Seebeck coefficient measures a material's ability to generate an electric voltage in response to a temperature gradient. The power factor ($PF = \sigma S^2$) indicates the effectiveness of a material in producing electricity from a gradient of temperature. Power factor is often used to evaluate TE materials when $\kappa$ is not accessible[18] or when net power generation prevails over efficiency[19]. The realm of TE materials operating at room temperature is dominated by the bismuth telluride ($Bi_2Te_3$) family. Despite their indisputable top performance ($ZT > 1$ )[20] and the recent advances towards the production of large-area, flexible, and printed $Bi_2Te_3$-based devices compatible with wearables [21–24], their widespread use is limited by their reliance on expensive elements with a relatively high environmental impact and geographically constrained availability.

Alternatively, organic thermoelectric (OTE) materials -often conducting polymers- are abundant, can be produced sustainably, are easy to process from solution, and are intrinsically soft. Those attributes make them a perfect alternative to $Bi_2Te_3$ for room temperature applications where performance is not the main factor to consider. Poly(3,4-ethylenedioxythiophene) polystyrene sulfonate (PEDOT:PSS) is one of the most popular thermoelectric polymers due to its superior ambient stability and relatively high performance[25]. PEDOT:PSS thin films hold the

[a.] Department of Materials Engineering, KU Leuven, Leuven, Belgium.
[b.] Department of Chemical Engineering, Soft Matter, Rheology and Technology (SMaRT), KU Leuven, Leuven, Belgium.
[c.] Department of Chemistry, KU Leuven, Leuven, Belgium.

record-high performance among organic TEs, boasting an impressive $ZT > 0.25$ and $PF > 500$ µW /(m K) [26,27]. The key to the high TE performance of PEDOT:PSS resides in its high electrical conductivity. PEDOT:PSS is composed of a conjugated conducting phase, PEDOT, and an insulating counter polyanion PSS. common commercial formulations of PEDOT:PSS include an excess of hydrophilic PSS in the blend that facilitates the dispersibility of the hydrophobic PEDOT phase in water. However this excess of PSS hinders the formation of a densely connected network of conductive PEDOT, and leads instead to isolated "pancake-like" PEDOT islands with low crystallinity that endow the material with low electrical conductivity (0.2-1 S cm$^{-1}$)[25]. One of the most typical ways to increase the electrical conductivity of pristine PEDOT:PSS is by adding the so-called secondary dopants, i.e. polar solvents such as dimethyl sulfoxide (DMSO), ethylene glycol (EG), etc., which promotes a morphology change from the "pancake-like" morphology to connected and highly crystalline PEDOT fibers[27,28]. The electrical conductivity can be further improved by removing the excess of insulating PSS via post-processing washing steps[29,30]. Recently, pre-processing has also been proposed to remove the excess PSS: vacuum-assisted filtration of a commercial PEDOT:PSS aqueous dispersion diluted in EG resulted in a high conductivity of around 2500 S cm$^{-1}$ [31]. Some secondary dopants for PEDOT:PSS, such as ionic liquids and non-ionic polymeric surfactants, like Triton X-100 or Zonyl, not only increase the electrical conductivity but also act as plasticizers to promote mechanical stretchability[32–35]. Other additives such as (3-Glycidyloxypropyl)Trimethoxysilane (GOPS)[36,37] and 4-dodecylbenzenesulfonic acid (DBSA)[38] enhance the mechanical properties and water stability by inducing cross-linking of the PSS or the PEDOT phase, respectively. Finally, Triton X-100 has also been reported to induce self-healing[39], making PEDOT:PSS an ideal material for wearables and soft electronics.

Attempts to build TE generators (TEG) with organic materials include almost exclusively devices with a planar geometrical configuration that does not adapt optimally to most real-life scenarios in which the hot and cold surfaces are opposed to each other and the heat travels perpendicular to the substrate surface[23]. The few 3D-structured OTE devices with high aspect ratio that have been reported involve filling pre-patterned wells[40] or painting pre-shaped legs[41]. 3D printing OTE materials remains elusive for the following reasons. Firstly, developing conducting polymer inks with the required rheology for DIW without jeopardizing their electrical performance is challenging. Recent efforts in the field demonstrated routes to formulate DIW-printable organic conductive inks with decent conductivities of tens of S/cm. Those routes include blending conjugated polymers with insulating phases, such as hydrogels or non-volatile liquids[42], and promoting electrostatic interaction between polymer chains[17,42–44]. The conductivity could be boosted further upon solvent drying, but substantial volume shrinkage and loss of shape fidelity occurs, thereby defeating the main purpose of 3D printing. Moreover, if the material is printed directly on a substrate, shrinkage during any stage of the process can lead to interfacial stress and potential delamination, complicating the direct integration of 3D-printed materials into devices[42]. Some authors propose freeze-drying or supercritical-drying of a PEDOT:PSS hydrogel as a successful route to avoid shrinkage of 3D-printed parts[45], resulting in porous and lightweight aerogels. Interestingly, aerogels present extremely low thermal conductivity, a desired trait for thermoelectrics applications[45–47]. However, this decrease in thermal conductivity comes at the expense of an even stronger reduction in electrical conductivity, resulting in an overall decrease in the figure of merit, $ZT$. A priori, this trade-off makes aerogels not interesting for TE devices. Moreover, although PEDOT:PSS aerogels can be formulated to be mechanically robust and flexible[48,49], both freeze- and supercritical-drying involve going through abrupt temperature differences, which complicates the integration of the aerogel with other materials due to potential thermal stress-induced delamination. Indeed, no demonstration exists yet of the direct integration of a PEDOT:PSS aerogel into a device.

In this study, we demonstrated omnidirectional DIW of highly conductive, lightweight, mechanically robust, and flexible PEDOT:PSS aerogel parts with high shape fidelity. Vacuum-assisted filtration was performed to remove the excess PSS from a PEDOT:PSS aqueous dispersion, which was subsequently concentrated to realize a paste with the right rheology for DIW. A judicious paste formulation based on Li salt and GOPS additives allows tailoring the balance between electrical conductivity and mechanical stretchability to adapt the material to particular applications. By devising a strategy for the direct integration of 3D-printed filaments on a stretchable substrate, we demonstrated 3D-structured stretchable interconnects and planar TEGs suitable for skin electronics. Moreover, we showed vertical (through-plane) printing of pillar-like OTE legs with a high aspect ratio of up to ~ 7, and a $ZT = 3.2 \cdot 10^{-3}$, comparable with the best reported OTE aerogels. The high porosity of the aerogel resulted in an ultra-low thermal conductivity of 0.065 W/(m K), which led to higher thermal gradients across the aerogel pillars than across dense pillars with comparable geometry. Surprisingly, the aerogel pillars also supplied more electrical power than their dense counterparts, despite presenting the aerogel pillars a lower $ZT$ and requiring ≈ 90% less material than the dense pillars. We show that this counterintuitive result originates from the unavoidable contact resistances present in real devices. The high TE performance of the developed PEDOT:PSS aerogel combined with the reduced footprint and high aspect ratio enabled by the 3D-printed pillars, yielded a high thermoelectric power generation of 26 nW/cm$^2$ under conditions mimicking a wearable device attached to the skin and operating indoor (ΔT = 15 K and no heat sink placed at the cold side). This areal power density value is the highest reported so far for skin-mounted OTEs harvesters and among the highest for OTEs in general. The lightweight and material reduction enabled by the aerogel phase also yielded a high gravimetric power density of 0.76 mW/kg (at ΔT = 15 K). This parameter, overlooked so far in the field, is paramount for the viability of portable and low-cost devices. The simultaneous high TE performance, mechanical flexibility and light weight of our 3D-printable OTE aerogels, along with their compatible integration on stretchable substrates, represent a new playground for the fabrication of unconventional stretchable interconnects and self-powered soft electronic systems for the next generation of soft robots and wearables.

## Results and discussion

PEDOT:PSS is one of the most high-performing and ambient-stable p-type organic thermoelectric materials, and these properties have been underpinned by extensive literature. Therefore, a dispersion of off-the-shelf high-conductivity PEDOT:PSS (Figure 1a) was used as benchmark material to prepare 3D-printable aerogel-based organic thermoelectrics compatible with a stretchable substrate. The process flow is illustrated in Figure 1b. The PEDOT:PSS dispersion was first mixed with DMSO for 3 days at 60 °C in a close vial. DMSO is a typical secondary dopant for PEDOT:PSS[25,34] that promotes a morphology transition from a low-conductivity structure consisting of PSS-isolated PEDOT islands to high-conducting and well-connected PEDOT fibers[27,50]. This morphology transition takes place in dispersion[27,50], and once it occurs the DMSO can be washed away to recover a greener water-based system (Figure 1c). DMSO must also be removed because it acts as an anti-freezing agent impeding the freeze-drying step planned downstream in the process flow. On the other hand, the PEDOT:PSS dispersion used in this study is known to contain 2.5 times more insulating PSS than conducting PEDOT, which hinders conductivity. The DMSO-to-water solvent exchange and the removal of excess PSS were achieved in a modified vacuum filtration system (Figure 1b) by filtering away the excess PSS from the PEDOT:PSS-DMSO mixture with the help of iterative water washing. The result was a PSS-rich DMSO/water solution to be discarded at the bottom of the setup, and a PEDOT-rich aqueous dispersion on top of the filter (Figure 1c). The efficient removal of excess PSS was confirmed by the UV-visible light spectrum of the filtered-out byproduct (Figure 1d), which displayed absorption features exclusively attributed to PSS (peak positioned ~ 200 nm[51]). In comparison, the original PEDOT:PSS dispersion showed both absorption features of PSS and PEDOT (band from 500 to 1200 nm[52]). Note that the spectra were normalized to the PSS feature. The PEDOT-rich dispersion remaining at the top of the filter was then transferred to an open Teflon container where it was partially dried at 60°C while continuously stirring until a hydrogel with the right viscosity for DIW was achieved. This corresponded to 1.42 g paste per 10 ml of the original PEDOT:PSS dispersion[53]. During this last concentration step, additives could be optionally introduced in the hydrogel to tune the mechanical characteristics of the final material (Figure 1b). In particular, we investigated bis(trifluoromethane)sulfonimide lithium salt (called Li salt in the rest of the paper) as a plasticizer to promote stretchability[33], and (3-Glycidyloxypropyl)trimethoxysilane (GOPS) as PSS crosslinker to strengthen the mechanical properties[54] (Figure 1c).

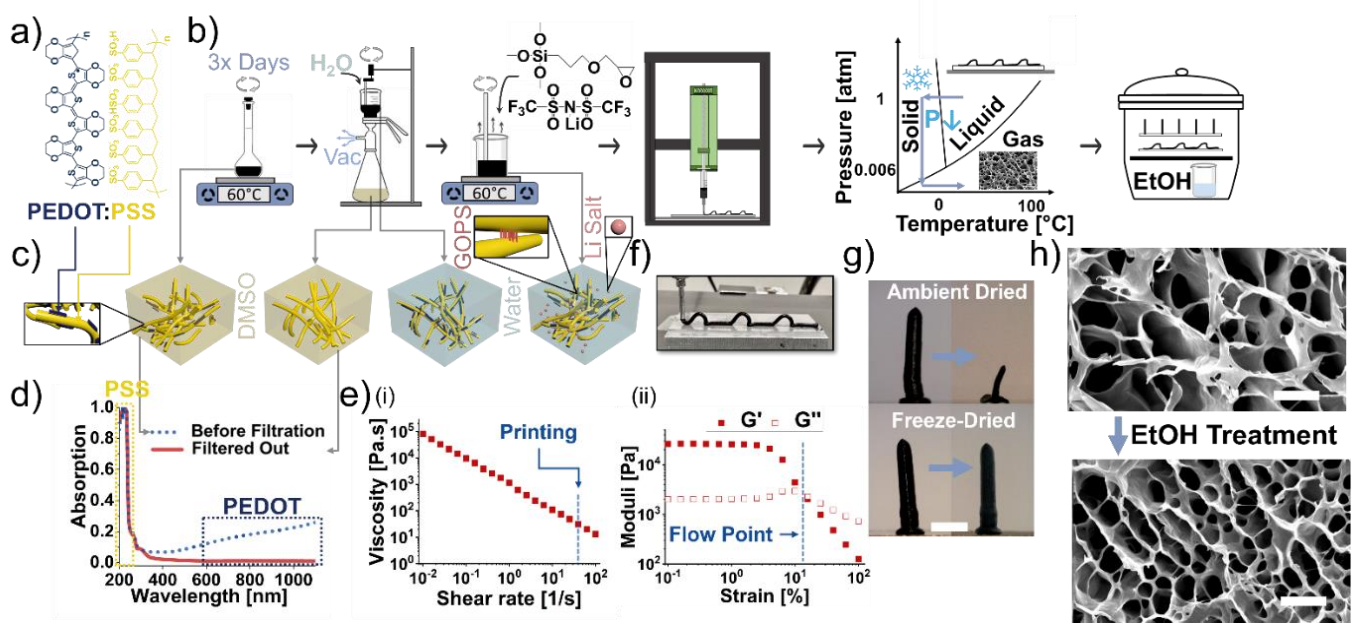

**Figure 1. Process flow of 3D-printed aerogel-based stretchable organic thermoelectrics.** a) Chemical structure of commercial PEDOT:PSS. b) Process flow of the paste formulation and direct ink writing (DIW) including: mixing PEDOT:PSS with DMSO secondary dopant, vacuum-assisted filtration of PSS excess and replacement of DMSO by water as dispersion medium (under continuous mixing), adding Li salt, concentrating and adding GOPS, DIW of the paste, freeze-drying, and post-treating with ethanol vapor. c) Schematic illustration of PEDOT:PSS and DMSO initial mixture, filtered out (discarded) solution of PSS in DMSO/water, PEDOT-rich PEDOT:PSS viscous dispersion in water, and PEDOT:PSS aqueous paste mixed with Li salt and GOPS. d) UV-Vis absorption spectrum of the PEDOT:PSS dispersion before filtration and the discarded filtered-out PSS-rich solution. e) Rheology of representative concentrated printable pastes including viscosity vs. shear rate flow curve (i) and strain sweep (ii). f) Optical photograph of representative 3D printed arches. g) Optical photograph displaying the shrinkage difference between ambient drying and freeze-drying for 3D-printed pillars (the scale bar corresponds to 5mm). h) Scanning electron micrographs of the concentrated aerogel made of filtered hydrogel before and after EtOH vapor post-treatment (the scale bars corresponds to 10 μm).

To verify whether the rheology of the hydrogels is suitable for DIW, flow curves; frequency, strain, and stress sweep tests; as well as creep-recovery tests (Figure 1h, S1) were conducted. The flow curves (viscosity versus shear rate, Figure S1a) confirm a desired low viscosity of ~ 50 Pa s at the printing shear rate, which is roughly estimated to be around 30 $s^{-1}$ (Eq. S1-S2). This low viscosity ensures flow in the nozzle. The flow curve data correspond to a yield stress plateau of ~ 1000 Pa (Figure S1b). This yield stress is more than sufficient to overcome retraction of the printed filament under the influence of surface tension, with the Laplace pressure being ~ 175 Pa for the used nozzle size (Eq. S5). Moreover, due to the high yield stress, the hydrogel can sustain the gravitational stress of extra deposited material on top, allowing the printing of vertical pillars up to an estimated height of 0.2 m with good shape retention (Eq. S6). The frequency sweep test shows the gel (elastic) character of the inks under low deformations (G' > G", Figure S1c). The strain sweep test (Figure 1h, S1d) demonstrates that the material exhibits dominantly elastic behavior up to a relatively large strain of around 10%, at which the flow point is reached. Interestingly, a creep-recovery test at a stress value below the yield stress unveiled a limited elastic recovery when the shear stress was removed (Figure S1e). This low elastic recovery is desirable to prevent shapes, which were extruded (under strain) from "bouncing" back to their original shape. Furthermore, significant viscous creep below yield stress was only observed for the reference formulation (hydrogel 0). Large viscous creep is to be avoided as it can lead to paste flow, which jeopardizes the shape fidelity of 3D-printed structures. Finally, our improved ink formulations contributed a higher yield stress value than the reference hydrogel (Figure S1b and S1f). The rheology study shows that the change in the morphology induced by additive or removal of excess PSS takes place in the dispersion. This hypothesis, which was supported by SAXS studies[27,50], is here corroborated by the rheology for the first time.

The rheology of the hydrogel allowed omnidirectional printing to form complex shapes such as out-of-plane arches (Figure 1f, Movie S1 and Movie S2) and high aspect ratio (~7) vertical pillars (Figure 1g) produced directly on a silicone stretchable substrate. This approach offers enhanced processing flexibility compared to conventional layer-by-layer deposition methods. However, similar to typical hydrogel printing processes, the printed material experiences dramatic and unpredictable shrinkage, which impairs both thermoelectric performance and shape retention, and leads to surface delamination[42]. To counteract the shrinkage, the printed materials were immersed in liquid nitrogen (LN) and subjected to freeze-drying to produce stable aerogels (Figure 1g). To reduce substrate delamination due to thermal stresses induced during freeze-drying, a thin film of semi-cured silicone was deposited on the substrate to serve as an adhesion layer. The porous aerogel structure (Figure 1h) offers advantages in terms of reduced material usage, lightweight, and low thermal conductivity, which are crucial for thermoelectric applications. However, the porosity also leads to a reduction in electrical conductivity. To partially mitigate this issue, the printed aerogels were subjected to ethanol vapor treatment. This post-treatment respected the original porosity of the material, improved its surface by welding loose edges (as depicted in Figure 1h and Figure S2), and led to a significant boost in electrical conductivity (Figures S3a and Figure S4a).

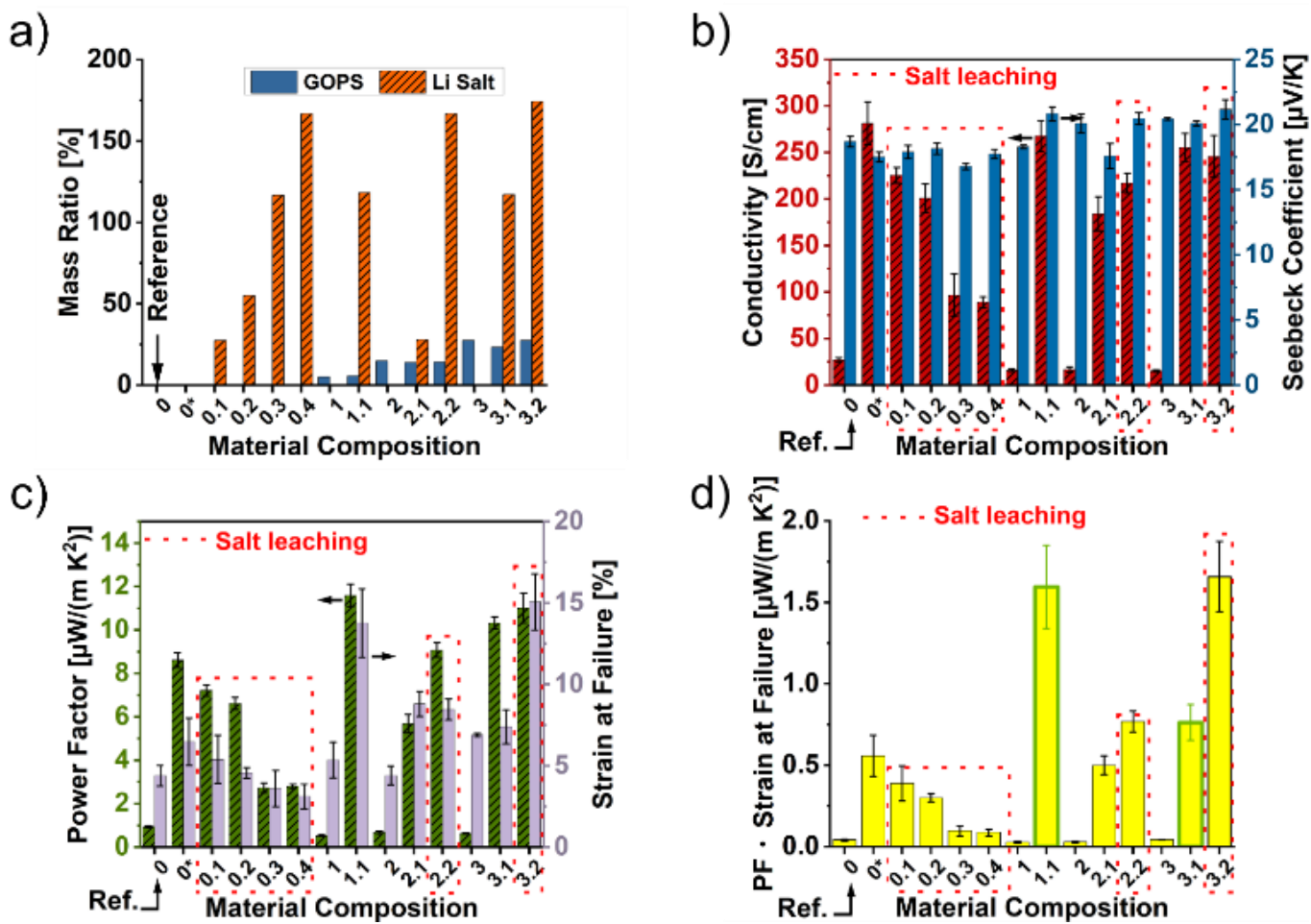

**Figure 2. Thick film thermoelectrical and mechanical characterization for material composition selection.** a) Mass ratio of Li salt and GOPS compared to solid PEDOT:PSS (assuming the solid content of PEDOT:PSS in the commercial dispersion is ~ 1.2 wt%) for each tested composition. The sample 0* refers to the benchmark PEDOT:PSS dispersion with 5%vol DMSO. b) Electrical conductivity and Seebeck coefficient. c) Power factor and strain at failure, and d) electromechanical figure of merit (power factor · strain at failure). All samples were post-treated via ethanol solvent annealing and characterized at ambient conditions.

To understand the influence of the addition of Li salt (plasticizer) and GOPS (crosslinker) on the mechanical and thermoelectric properties of PEDOT:PSS thick films, different formulations were developed, as depicted in Figure 2a and Table S1. Pristine PEDOT:PSS without additives (formulation 0) and the widely-reported high-conducting PEDOT:PSS with 5%vol DMSO[25,34] (formulation 0*) were taken as references. All samples were subjected to ethanol vapor annealing post-treatment, which significantly enhanced the power factor by boosting markedly the electrical conductivity and to lesser extent the Seebeck coefficient (Figure S3 and Figure S4). Vapor annealing with a polar solvent has been reported as an alternative strategy to realize secondary doping leading to an improved conductivity[55]. Figure 2b and Figure S4 shows the room-temperature electrical conductivity and Seebeck coefficient of the post-treated films made from the different formulations. While material composition showed only a moderate effect on the Seebeck coefficient (ranging from 18 – 21 µV/K), the electrical conductivity increased dramatically with the addition of small amounts of Li salt from 27 ± 4 S/cm for pristine PEDOT:PSS (sample 0) to 225 – 275 S/cm for samples 0.1, 1.1 and 3.1. These high electrical conductivity values are comparable to those measured for the high-conductivity benchmark PEDOT:PSS with 5%vol DMSO. However, excessive Li salt content resulted in the salt leaching out of the material (marked with the red dotted frames on the graphs), which led to instability and eventual conductivity degradation. The statistical significance of Li salt addition and leaching in the conductivity has been confirmed in Figure S4. Interestingly, the addition of GOPS alone showed little influence on the electrical conductivity. Yet, it promoted the assimilation of Li salt in the PEDOT:PSS, pushing up the maximum stable loading without leaching. This synergy between Li salt and GOPS for ethanol vapor post-treated films was also statistically confirmed (Figure S4), and resulted in a maximum power factor of 11.6 ± 0.5 µW/(m $K^2$) for the stable composition 1.1. Regarding mechanical performance, in the absence of GOPS the Young's modulus values decreased with Li salt, while the strain at failure increased (samples 0 and 0.1 to 0.4 in Figure 2c, Figure S3b, and Figure S4d). Adding GOPS alone (samples 1, 2, 3) showed no clear trend in Young's modulus but it also increased slightly the strain at failure. Hence, similar to what we observed for the electrical properties, the addition of both GOPS and Li salt contributed to boosting mechanical properties. In particular, we measured an increase of strain at failure from 4.4 ± 0.6 % for pristine PEDOT:PSS (sample 0) to 14 ± 2 % for sample 1.1.

However, it is worth noting that contrary to what we observed for the power factor, the positive effect of additives on the mechanical properties of the material seems to be independent, with no statistically significant synergy between additives being identified (Figure S4). To reflect the electromechanical benefits of Li salt and GOPS we propose a new combined property, the electromechanical figure of merit (PF · strain at failure) (Figure 2d). According to this electromechanical figure of merit and the samples' leaching stability, formulations 1.1 and 3.1 were selected as the most promising materials to make aerogels.

The electrical conductivity and the Seebeck coefficient of the selected aerogel compositions based on optimized power factor and stretchability (samples 1.1 and 3.1) and a reference aerogel made of pristine PEDOT:PSS (sample 0) are shown in Figure 3a. When compared with their thick film counterparts, the aerogel samples displayed the same trends across the different material compositions (Figure S5). They also showed a similar Seebeck coefficient (only slightly higher) but about 20 times lower electrical conductivity only 16 ± 2 S/cm for the most conducting aerogel formulation 1.1. This substantial decrease in electrical conductivity was expected considering the large amount of air present in the aerogel (~ 90%vol). In an attempt to further increase electrical conductivity (and power factor) of the aerogels, the initial PEDOT:PSS aqueous dispersion was diluted in DMSO and subjected to the filtration process described in Figure 1, before freeze-drying. The filtration aimed to promote highly-conducting PEDOT fibrous morphology[27,50] and remove the excess of insulating PSS from the paste. The filtration process was successful in boosting the electrical conductivity of pristine PEDOT:PSS aerogels from 2.3 ± 0.2 S/cm to 13.4 ± 0.3 S/cm (sample 0f). However, filtration did not improve the performance of those formulations containing additives (Figure S5). This is why the filtered versions of 1.1 and 3.1 (1.1f and 3.1f respectively) were not considered for further study. In the end, power factors as high as 0.89 µW/(m K$^2$) and 0.69 µW/(m K$^2$) were achieved for aerogels 1.1 and 0f, respectively (Figure 3b). The evolution of the aerogels electrical conductivity and Seebeck coefficient with temperature (from -10 °C to 90 °C ) was also investigated and it is shown in Figure S6. Interestingly, conductivity increased with temperature for sample 0, but was flat for 0f and decreased for 1.1. and 3.1., suggesting a transition from a charge hopping-dominated transport for 0f, to metallic-like transport for 1.1 and 3.1[56]. The Seebeck coefficient followed a characteristic u-shaped curve with the minimum inconveniently located around temperature. To the best of the authors' knowledge, this u-shaped behavior has never been reported.

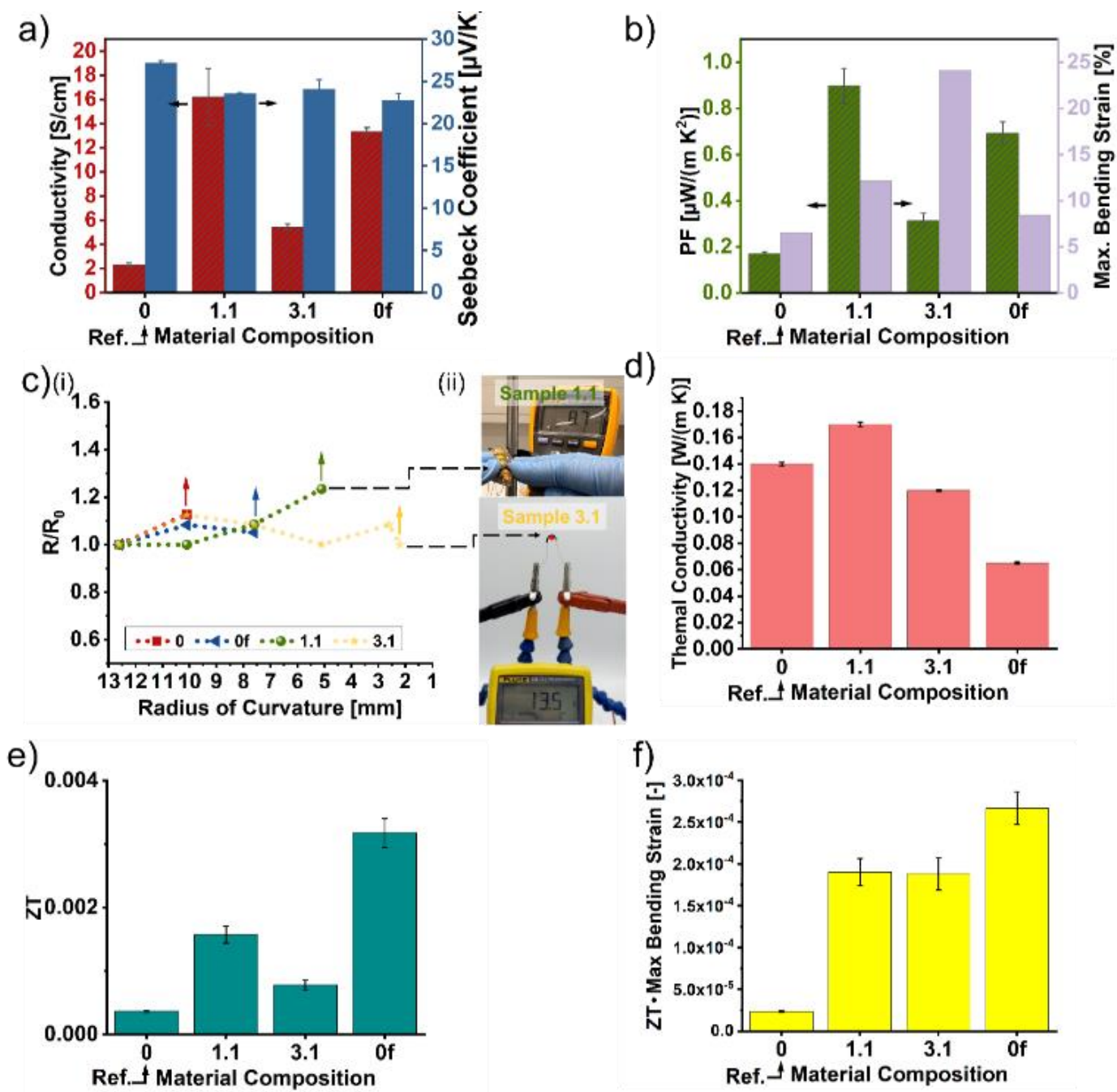

**Figure 3. Thermoelectrical and mechanical characterization of 3D printed aerogel filaments for material selection.** a) Electrical conductivity and Seebeck coefficient. b) Power factor and maximum survived bending strain. c) The electrical resistance changes of printed aerogels upon bending around a different radius of curvature (i) and optical photograph of the utilized setup (ii). d) Thermal conductivity. e) Figure of merit (ZT). f) ZT · maximum bending strain. All samples were post-treated via ethanol solvent annealing and characterized at ambient conditions. (0, 0f, 1.1 and 3.1 refer to reference PEDOT:PSS, filtered PEDOT:PSS, and compositions with Li salt plus low and high concentration of GOPS, respectively, as described in Figure 2a).

The aerogels were intended to be printed as deformable 3D structures to be integrated into stretchable devices. The enhanced stretchability provided by the combination of Li salt and GOPS in thick films was thus exploited to endow thick filaments of printed aerogels with mechanical bendability (Figure 3b,c and Movie S3). Considering that formulation 1.1 and 3.1 had the highest strain at failure in thick films (Figure 2c), it is not surprising that we found their aerogel versions surviving also the highest bending strain (12.1% and 24.1%, respectively). However it should be noted that unlike in the case of dense films, the aerogel version of 3.1 is more stretchable than 1.1. This difference could be attributed to the influence of microstructure in the bending strain. Indeed, aerogel 3.1 presented larger pores and lower density (more %vol of air) than aerogel 1.1, which presumably allowed superior accommodation of strain and led to enhanced filament bendability (Figure S2 and Table S2). Conveniently, the filtration process not only improved the power factor of the pristine PEDOT:PSS, but it also its flexibility (increasing its maximum bending strain from 6.5% to 8.4%), presumably due to the removal of loose PSS leading to an increased physical interaction between PEDOT fibers.

Furthermore, the filtration step reduced the already-low thermal conductivity of the PEDOT:PSS aerogel 0 (κ = 0.1397 ± 0.0013 W/(m K)) to an extremely low value of κ = 0.0651 ± 0.0008 W/(m K). This observation suggests that PSS is strongly involved in heat conduction. Indeed aerogel 0f showed a lower thermal diffusivity than 0. Note that similar density, ρ, and specific heat capacity, $C_p$, values were measured for both aerogels (Table S2) and that κ = α·ρ·$C_p$. Such low thermal conductivity of aerogel 0f led to the highest figure of merit ZT = T·PF/κ = (3.2 ± 0.2) 10$^{-3}$ of all the aerogel formulations tested (Figure 3e). This value is among the highest reported for OTE aerogels (Table S5). The samples with additives also displayed a lower thermal diffusivity, α, than the reference but it was experimentally challenging to fabricate them with the same low density, which led to rather equal thermal conductivity (Table S2).

Our methodology allows for creating a library of DIW aerogels with different thermoelectric and mechanical properties. Since a tradeoff between thermoelectric and mechanical properties was found, a new thermoelectric-mechanical figure of merit for aerogel filaments (*ZT* · maximum bending strain) is introduced in Figure 3f to screen the overall best compositions for our library. The ratio (*ZT* · maximum bending strain) was comparable for all three processed samples, but significantly higher than for the reference material (aerogel 0), which proved the benefits of our processing procedure.

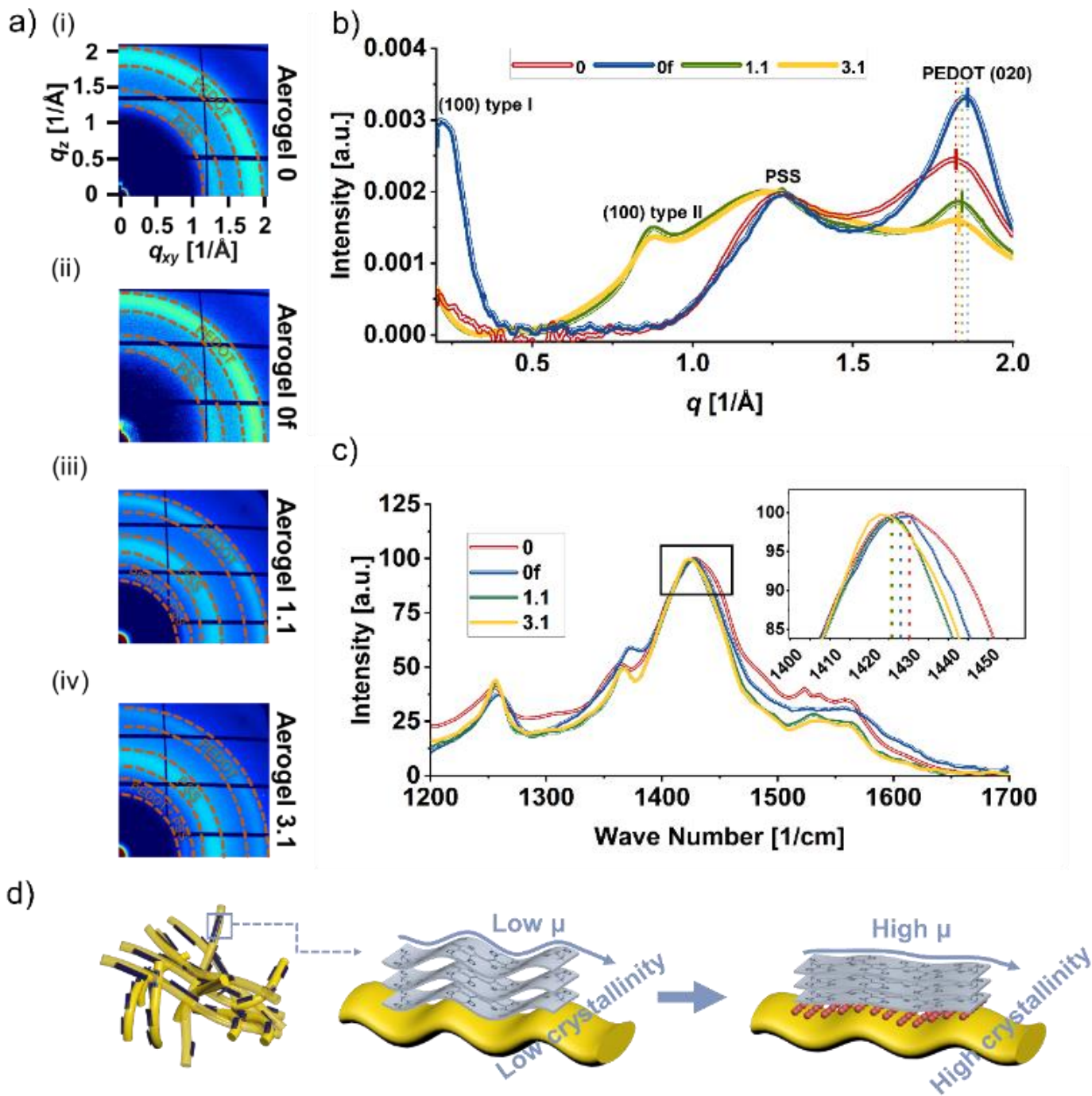

**Figure 4. Material characterization of the aerogels made from the selected compositions.** a) Synchrotron wide angle X-ray scattering (WAXS) 2D patterns of aerogels 0 (i), 0f (ii), 1.1 (iii) and 3.1 (iv). b) WAX line-cut intensity profile normalized to PSS peak intensity. c) Raman spectra and close view inset of the main peak shift. d) Schematic illustration of morphology change in PEDOT:PSS after the addition of Li salt ($\mu$ = mobility). (0, 0f, 1.1 and 3.1 refer to reference PEDOT:PSS, filtered PEDOT:PSS, and compositions with Li salt plus low and high concentration of GOPS, respectively, as described in Figure 2a).

To understand how the utilized additives boost electrical transport, we examine the morphology of the different printed aerogels using synchrotron wide-angle x-ray scattering (WAXS) and Raman spectroscopy.

2D WAXS patterns are depicted in Figure 4a, and their 1D line-cut intensity profiles are presented in Figure 4b. The notable peaks observed around 1.2 Å$^{-1}$ and 1.8 Å$^{-1}$ across all samples are attributed to the PSS halo and the PEDOT π–π stacking distance (020), respectively[29]. All our processed aerogels showed a shift of the (020) peak to $q$ values higher than for the pristine PEDOT:PSS aerogel (0) used as a reference, suggesting a decrease in the π–π stacking distance. There is a direct link between increased electrical conductivity and the denser crystal packing resulting from a shorter π–π stacking distance. Closer molecules promote overlapping of the π orbital between neighboring molecules, thereby facilitating smoother interchain transport and ultimately increasing electrical conductivity[57,58]. This correlation was confirmed for all our formulations, which displayed higher conductivities than the pristine material, and was interestingly linear (Figure S7a). In contrast, only a weak positive correlation between the π–π stacking distance and the Seebeck coefficient was observed (Figure S7b). The aerogel 0f, from which the excess PSS was removed, displayed a comparatively large (020) peak. This can be attributed to the fact that the intensity of all the curves was normalized to the PSS peak (at ~ 1.2 Å$^{-1}$). However, this sample also displayed the strongest (020) shift, accompanied by an intense (100) type I peak attributed to the PEDOT:PSS lamellar structure[58]. Both findings evidenced the strong beneficial effect in crystal ordering associated with the removal of excess PSS, which underpins the boost in conductivity observed for 0f. On the other hand, for those samples with additives (1.1 and 3.1), the emergence of a new lamella peak, the (100) type II (at ~ 0.88 Å$^{-1}$), was observed. The type (100) type II peak has been previously reported for PEDOT:PSS doped with ionic liquids and was ascribed to a different PEDOT:PSS lamellar stacking in which adjacent PEDOT chains are staggered and closer to each other[33,58]. To verify that the origin of this new peak arises from the interaction between PEDOT and Li salt rather than from potential Li salt aggregates, we examined PSS films blended with Li salt with Grazing Incidence WAXS (GIWAXS). No diffraction was detected at 0.88 Å$^{-1}$ for the bare PSS + Li salt film (Figure S7c). Compared with type I, type II layer packing has been thought to facilitate charge transport between adjacent PEDOT chains, which results in improved electron conductivity[58]. According to this analysis, both the reduction in π–π stacking distance and the appearance of the type II lamella stacking can explain the improved electrical conductivity upon Li salt addition. Moreover, the increase in the PEDOT phase packing observed upon addition of Li salt suggests that the Li salt molecules primarily occupy less ordered/amorphous regions, thereby softening them. This results in a morphology consisting on a crystalline nanofiber network surrounded in a soft matrix, which enhances the stretchability of the structure[33].

The Raman spectra in Figure 4c reflect PEDOT chain's conformation and oxidation state. The primary Raman peak, occurring between 1400 and 1500 cm$^{-1}$, corresponds to the stretching vibration of $C_\alpha=C_\beta$ on the five-member PEDOT ring[59]. The presence of additives in our aerogels resulted in a red shift of this Raman peak[33,58]. Specifically, the peak position for aerogel 0 was at 1430 cm$^{-1}$, while for aerogels 0f, 1.1, and 3.1, it was at 1428, 1426 and 1426 cm$^{-1}$, respectively. This red shift trend matched the reported literature on PEDOT:PSS doped with secondary dopants or ionic species[33,58,59]. In contrast, the addition of GOPS alone was shown to not affect the Raman peak shift[60]. Hence, we can conclude that the Li salt is the main responsible for the red shift of the peak signal, suggesting a conformational change in PEDOT from its original benzoid structure (associated with a coiled PEDOT backbone conformation) to a quinoid structure (associated with linear PEDOT backbone conformation). This occurs presumably due to the Li salt contributing to partially screening the strong coulombic interaction between PEDOT and coiled PSS[27,33] (Figure 4d). An extended backbone enhanced the planarity of PEDOT and facilitated the stacking of PEDOT chains, thereby promoting charge delocalization and efficient charge transport. This result confirmed the conclusions retrieved from the WAXS characterization.

The screening of the aerogel compositions in terms of thermoelectric and mechanical properties provided us with a material library to consult when a material is sought for specific applications. This library combined with the shape freedom enabled by 3D printing, results in a powerful toolbox for innovative applications involving 3D soft electronics. To illustrate this concept, we developed three demos where the material properties matched a particular application: From the three processed aerogels, 3.1 presents the best bendability but the lowest *ZT*. Hence, aerogel 3.1 would be the most suitable material for applications where mechanical deformation prevails over TE performance, such as in stretchable interconnects (demo #1). Aerogel 1.1 offers the best balance between TE performance and bendability, thereby fitting best the requirements of stretchable thermoelectrics (demo #2). Finally, aerogel 0f stands out for its superior *ZT* at the expense of modest mechanical stretchability. This makes 0f the best option for TE applications that do not require much mechanical strain (demo #3).

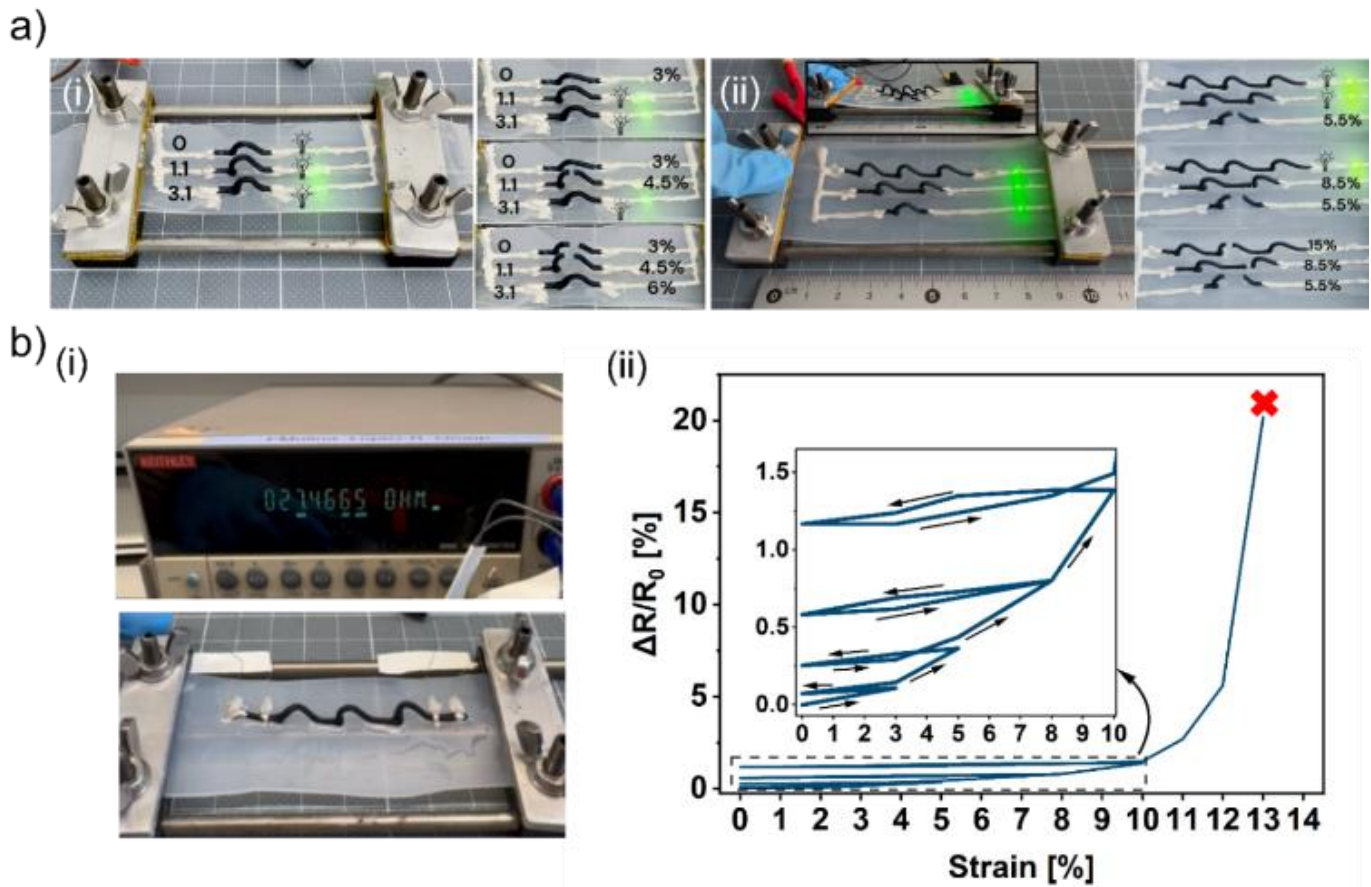

**Figure 5. Demo #1: Stretchable interconnects based on through-plane arches of aerogel 3D-printed on an elastomer substrate.** a) Stretchability test for different material compositions (i), and for different device designs (ii). b) Optical photograph of the stretching setup used with a demo consisting of a triple arch single line (aerogel 3.1) mounted in it (i), and the resistance change versus iterative applied strain (ii).

Figure 5a(i) depicts the experiment designed to confirm that 3.1 is the most suitable aerogel for demo #1. We produced stretchable lines from our bendable aerogels by printing them as out-of-plane arches directly on an elastomeric substrate. The device's stretchability could be tuned by the material composition. To proof this point, circuits composed of a single aerogel arch and stretchable Ag paste connecting to a small LED were printed on a stretchable substrate. Upon strain application, the circuit containing the reference aerogel (0) failed at 3% strain (the LED switched off), while the circuits made with aerogels 1.1 and 3.1 withstood 4.5% and 6% strain, respectively (Movie S4). Besides the material composition, the 3D shaping freedom offered by DIW could be exploited to tune stretchability. To illustrate this concept, we chose the most flexible material from our library (3.1). Note that the relatively low thermoelectric performance of 3.1 is not a bottleneck for this application. The aerogel was printed in three different circuits, each with a different number of arches (Figure 5a(ii)). In principle, the higher the number of arches, the larger the strain at circuit failure because the arches deformation help accommodating strain. As expected, the single-arched line failed at 5.5% strain, followed by the double- and triple-arched lines that failed at 8.5% and 15% strain, respectively (see Movie S5). The mechanical stability of the triple-arched line was tested via iterative stretching to increasing strains (Figure 5b and Movie S6). This test shows that the device's resistance change is very low, below 1.5 %, up to the onset of failure ~ 10% strain. A mild hysteresis was observed in the contraction phase after stretching beyond the previous maximum strain value. The reason for this behavior is likely irreversible plastic deformation of the material upon stretching.

The 3D-printed stretchable lines shown above can be repurposed as the p-type leg of a stretchable thermocouple able to harvest waste heat, responding in this way to the standing challenge of stretchable electronics of finding mechanically complying power sources. For this second demo (demo #2), we selected aerogel 1.1 for its balance between mechanical properties and thermoelectric performance. We completed the thermocouple with stretchable silver acting as the n-type leg (although it is in reality a poor p-type material with a Seebeck coefficient of $S_n$ = +6.5 µV/K) (see Figure 6a and Movie S7). As depicted in Figure 6b, the thermocouple showed a typical thermoelectric generator behavior with the output power increasing quadratically with the temperature difference across the TE legs, $\Delta T_L$, (Figure 6c) according to the equation[41]:

$$P_{max} = \frac{[(S_p - S_n)\Delta T_L]^2}{4R_i} \qquad (Eq. 1)$$

where $S_p$ and $S_n$ are the Seebeck coefficient of the PEDOT:PSS aerogel 1.1 and the stretchable silver, respectively, and $R_i$ is the internal resistance of the thermocouple (detailed explanation in Supplementary Info, section 5). A maximum output power of ~ 0.95 nW was measured for a temperature difference of 18 °C. The thermocouple, subjected to a temperature difference of 10 °C, could be stretched up to 13% strain without performance loss (see Figure 6d). This strain invariability can also be understood from a theoretical standpoint (see Eq. S8-S11 and corresponding analysis). The versatility of DIW permitted to change the layout of the PEDOT:PSS freely. To illustrate this, another demo with similar performance was developed in which the PEDOT:PSS leg was shaped as a planar serpentine structure instead of as out-of-plane arches (see Figure S8 and Movie S8).

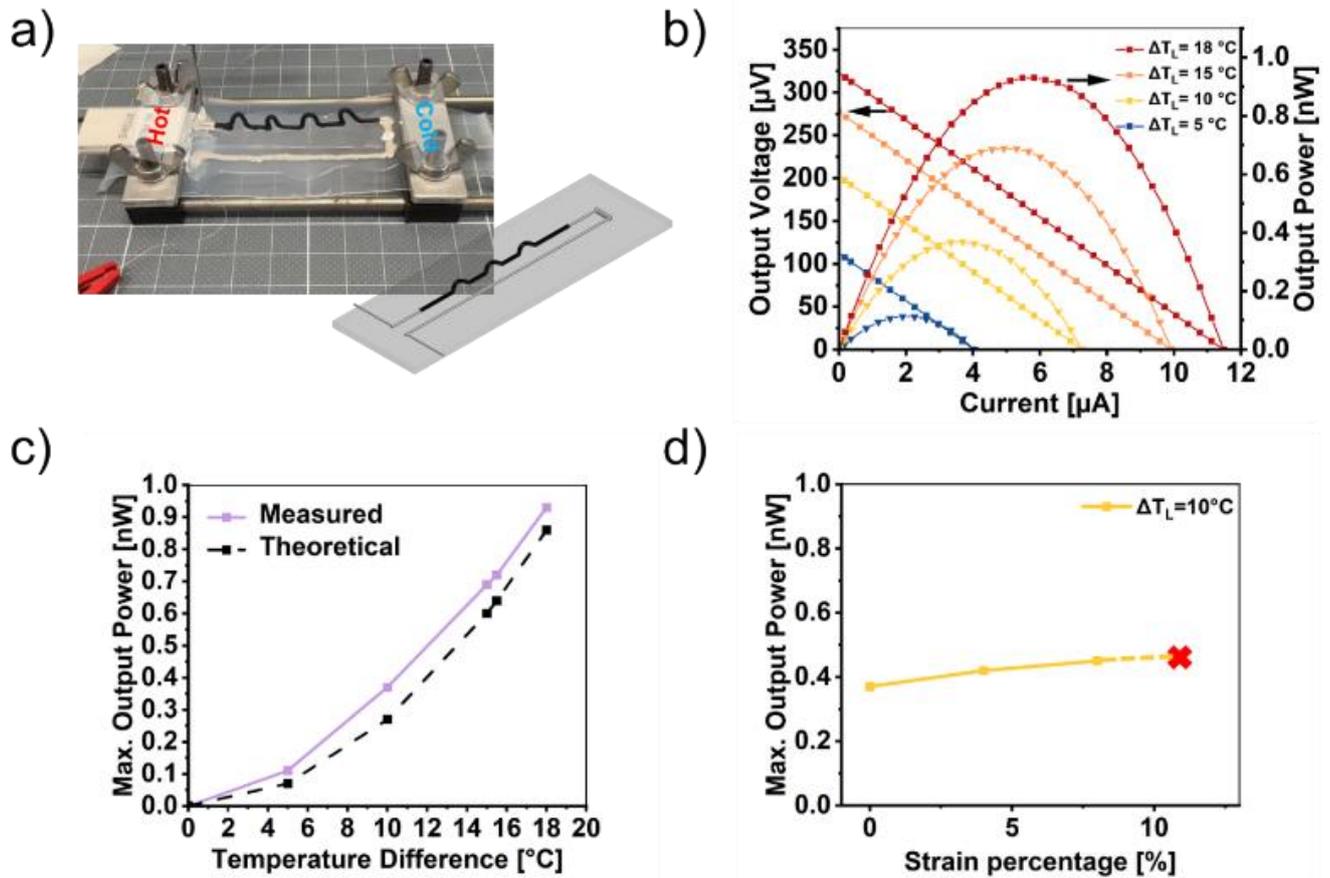

**Figure 6. Demo #2: Stretchable planar thermoelectric generators based on 3D-printed through-plane arches.** a) Optical photograph of a planar stretchable thermocouple composed of out-of-plane PEDOT:PSS (material composition 1.1) arches and painted stretchable silver. The device is mounted in a house-made stretching set-up and coupled to a Peltier unit at one end to generate a thermal gradient. b) Output voltage and output power vs. output current for different resistive loads and temperature difference across the device. c) Maximum output power at different temperature differences across the device. d) Maximum output power vs. applied strain for a 10°C degree temperature difference across the device.

Aerogels could be suitable for thermoelectric applications because their low thermal conductivity, resulting from their porous structure, can enhance the figure of merit *ZT*. However, this same porous structure lowers the electrical conductivity even further than the thermal conductivity (21.5-fold versus 8.5-fold for our materials), resulting in a ZT for aerogels almost 2.5 times lower than that of their denser counterparts. While this situation might discourage the use of aerogel for thermoelectrics, in real-world applications the electrical and thermal transport is heavily influenced by unavoidable contact resistances, which can alter the relative importance of *ZT*. Contact resistances strongly harm the performance of TEG and should be minimized, but they are specially difficult to eliminate in small devices like the one proposed in this study. A detailed analysis of the maximum output power provided by a TE leg, $P_{max}$, with electrical and thermal contact resistances, $R_C$ and ($R'_H + R'_C$) respectively, taken into consideration, leads to the following variation of Eq. 1 (see Eq. S12-S18, Figure S9, and corresponding analysis):

$$P_{max} = \frac{(S\,\Delta T_L)^2}{4R_i} = \frac{(S\,\Delta T_E)^2 \left[\frac{\frac{L}{A}}{\frac{L}{A} + \kappa\,(R'_H + R'_C)}\right]^2}{4\left(\frac{1}{\sigma}\frac{L}{A} + R_C\right)} \quad (Eq.\,2)$$

Where $\Delta T_E$ is the temperature difference applied externally to the device and $\Delta T_L$ is the actual temperature difference across the TE leg; $R_i$ is the internal electrical resistance of the device, composed of the net contribution from the TE leg and the electrical contact resistance $R_C$; $R'_H$ and $R'_C$ are the thermal contact resistance for hot and cold end, respectively; $L$ and $A$ are the length and cross-sectional area of the TE leg; $S$ is the Seebeck coefficient; and $\sigma$ and $\kappa$ are the electrical and thermal conductivity of the material, respectively. According to Eq. 2, in devices presenting high electrical and thermal contact resistance, a material with high $\sigma$ and $\kappa$ would produce less power than a counterpart with low $\sigma$ and $\kappa$, even if the former has a comparatively higher $ZT$. We illustrate this situation in Figure S10 for different materials, including those comparable to our dense 0* and aerogel 0f formulations, subjected to idealistic and realistic $R_C$ and ($R'_C + R'_H$) scenarios. The reason for this counterintuitive behavior is twofold: On one hand, the benefit of high $\sigma$ is masked by a dominant electrical contact resistance; on the other hand, a low $\kappa$ leads to a more significant temperature difference across the TE leg, $\Delta T_L$, for the same externally applied temperature difference, $\Delta T_E$, (Figure S11) and the larger the thermal contact resistance, the more marked this behavior is. The latter aspect was already highlighted in the seminal paper by Suarez *et al.* within the context of skin-mounted TEGs, in which a large thermal contact resistance exists between the human skin and the hot end of the TEG, and between the cold end of the TEG and the surrounding air[61]. Furthermore, the benefit of a lower κ is magnified in wearables and other applications that cannot afford to use bulky heat sinks on the cold side to reduce the thermal contact resistance with the air.

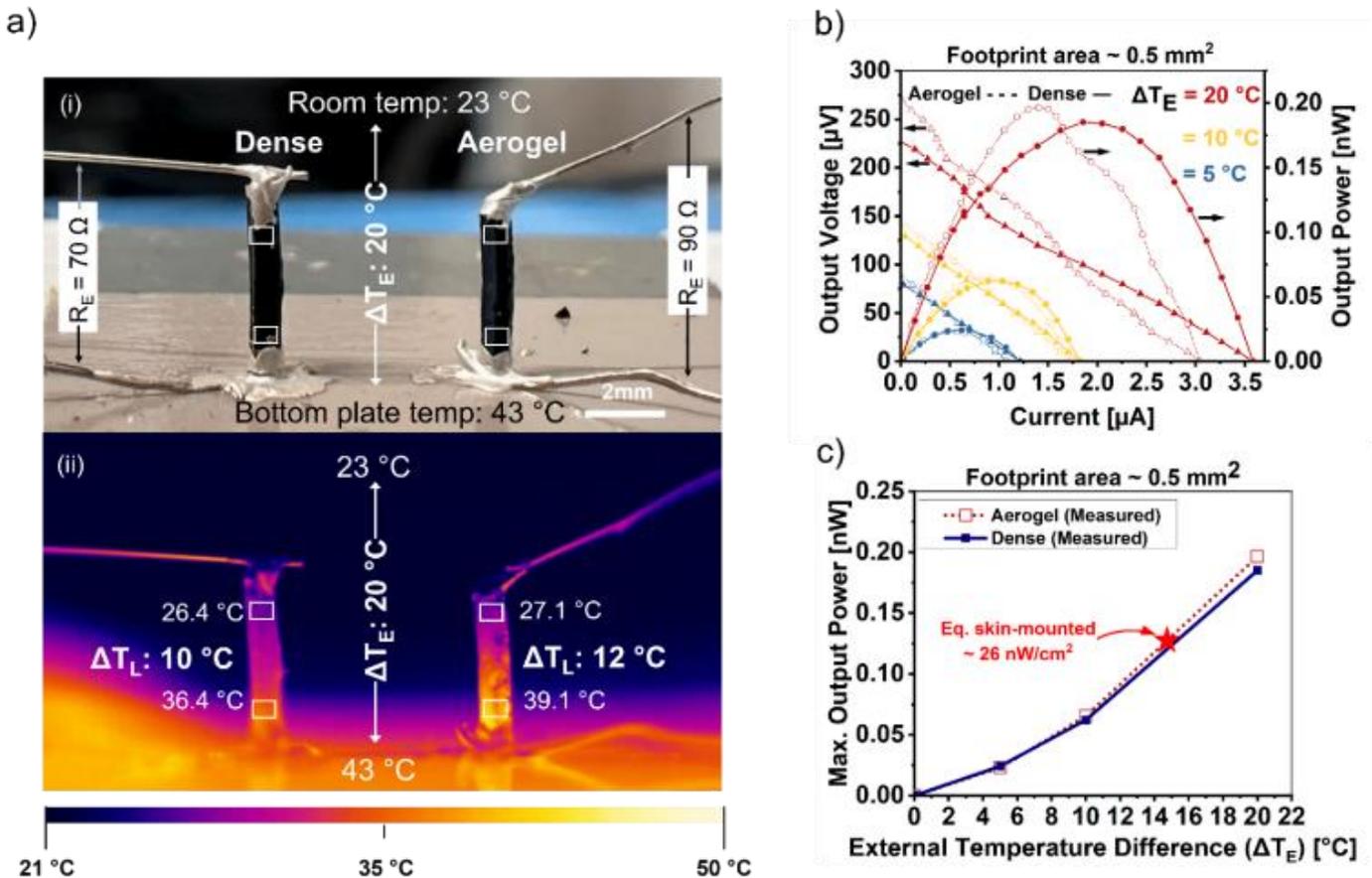

**Figure 7. Demo #3: Thermoelectric performance of 3D printed pillars. Comparison between dense material and aerogel.** a) Optical image of the two pillars side-by-side showing the electrical connections. The applied external temperature across the device ($\Delta T_E$), as well as the measured total internal electrical resistance of the device are indicated (i), representative frame obtained with a thermal camera displaying the actual temperature difference across the TE leg ($\Delta T_L$) (ii). b) Output voltage and output power change vs. output current for different resistive loads and temperature difference across the device . c) Maximum output power comparison at different external temperature differences across the device.

The previous analysis suggests the promise of using aerogels in TEG. To confirm this point experimentally, we printed a pillar made of the dense 0* PEDOT:PSS next to another pillar made of the aerogel 0f, both with similar dimensions (demo #3 depicted in Figure 7a(i)). To reach the same diameter for both dry pillars, the 0* pillar was originally printed with a larger diameter to account for its shrinkage upon drying. The formulation 0f was used in demo #3 because it is the material from our library with the best ZT, and its poor mechanical performance is not an issue in this scenario. To mimic a realistic skin electronics scenario, the pillars were welded to a thermally conducting soft substrate using Ag paste and were electrically interconnected at the top and bottom using the same Ag paste. The measured internal resistances for dense and aerogel interconnected pillars were 70 Ω and 90 Ω, respectively. Since the calculated net material resistance (1/σ L/A) for the aerogel and the dense legs is only 4.8 Ω and 0.2 Ω, respectively, the contact resistance dominated the internal resistance. The demo #3 was placed on a hotplate to heat the bottom end of the pillars while the top was left free to cool down by normal air convection/conduction. The external temperature ($\Delta T_E$) was measured using a thermometer and a thermocouple, and the temperature difference across the leg ($\Delta T_L$) was monitored with a thermal camera, as illustrated in Figure 7a(ii). It could be appreciated in the thermal images that, as predicted by Eq. 2, $\Delta T_L$ was larger for the aerogel than for the dense pillar, even though both pillars were subjected to the same $\Delta T_E$. The curves in Figure 7b display the typical behavior of TE generators for both pillars, where the output voltage and output power increased linearly and quadratically (Figure 7c), respectively, with $\Delta T_E$. Interestingly, despite the aerogel's ZT being less than half the ZT of the dense material (2.4 $10^{-3}$ vs 6.0 $10^{-3}$), the aerogel pillar could produce more output power than the dense pillar, because it could hold a larger temperature gradient across itself.

An areal power density of 26 nW/cm$^2$ was achieved for the aerogel at $\Delta T_E$ = 15 °C, mimicking the conditions of indoor skin electronics (substrate at 37 °C and air at 22 °C). This power density value is comparable to the best reported value for organic TEs in the existing literature (Table S5) [40,62] despite requiring 10 times less material. When computing the gravimetric power density, a merit that we expect to become increasingly relevant for wearables and small autonomous systems, the aerogel reached a value of 7.6 $10^{-4}$ W/kg (Ag paste is ignored in the calculation).

Our results support the potential of aerogels for out-of-plane thermoelectric devices, especially for wearables and many other cases requiring low power and in which miniaturization makes it challenging to reduce electrical contact resistance, the thermal interface between the device and the hot surface cannot be not optimized, or the lack of a heat sink at the cold side leads to a high thermal contact resistance at the cold side.

## Conclusions

In this study, we showed the possibility of omnidirectional 3D printing of PEDOT:PSS aerogels with tunable mechanical and electrical performance achieved via different paste formulations. We developed a process flow that allowed us to 3D print by direct ink writing aerogels directly on top of a stretchable substrate with sufficient attachment to make unconventional stretchable interconnects and thermoelectric devices. The devices' mechanical and thermoelectric performance could be tuned to adapt to different applications by: 1) the selection of proper material composition, and 2) the freedom of 3D printing design. We illustrated this versatility through three demonstrations: stretchable interconnects (demo #1), a stretchable planar thermoelectric generator (demo #2), and a vertical thermoelectric generator (demo #3). A comparison between the thermoelectric performance of a dense and an aerogel pillar revealed that utilizing aerogels is advantageous for energy harvesting in the presence of contact resistances.

## Author Contributions

Author contributions are defined based on the CRediT (Contributor Roles Taxonomy) and listed alphabetically. Conceptualization: F.M.L., H.E.B.. Data curation: F.M.L., H.E.B.. Formal analysis: F.M.L., H.E.B., T.Y., V.N., S.D.S., B.Z., H.X., R.C.. Funding acquisition: F.M.L., R.C.. Investigation: F.M.L., H.E.B., T.Y., V.N., S.D.S., D.H., B.Z., H.X.. Methodology: F.M.L., H.E.B., T.Y., V.N., S.D.S., B.Z., H.X., I.F., M.R., R.C. Project administration: F.M.L.. Supervision: F.M.L., M.R., R.C.. Validation: F.M.L., H.E.B., T.Y., V.N., S.D.S., B.Z., H.X., M.R., R.C.. Visualization: F.M.L., H.E.B.. Writing – original draft: F.M.L., H.E.B.. Writing – review and editing: F.M.L., H.E.B., T.Y., V.N., S.D.S., D.H., B.Z., H.X., I.F., M.R., R.C..

## Conflicts of interest

There are no conflicts to declare.

## Acknowledgements

This project has received funding from the European Research Council (ERC) under the European Union's Horizon 2020 research and innovation programme (grant agreement n° 948922) 3DALIGN (H.E.B., T.Y., B.Z. and F.M.L.); V.N. is a Fundamental Research Ph.D. fellow at FWO and his work was supported by the Research Foundation – Flanders (project number 11K9223N). S.D.S., R.C.: This project has received funding from the European Research Council (ERC) under the European Union's Horizon 2020 research and innovation programme: Grant Agreement No. 948739 – PEM-Sprint. D.H. was funded by Erasmus+ mobility programme (2021-1-TR01-KA131-HED-000003626). I.F. acknowledges the support of the Internal Funds KU Leuven, C1 project C14/21/078; H.X.: The project (40007563 CONNECT) has received funding from the FWO (project G0J4922N).

The authors acknowledge Prof. Bart Goderis (KU Leuven) and the DUBBLE (Dual Belgian Beamlines) at The European synchrotron ESRF for access to the beamline BM26 to perform WAXS. We would like to express our sincere gratitude to Joop van Deursen, Iris Cuppens, Jelle De Ceulaer, and Danny Winant for their support and assistance throughout this research endeavour.

**Supplementary Information**

**Omnidirectional 3D printing of PEDOT:PSS aerogels with tunable electromechanical performance: a playground for unconventional stretchable interconnects and thermoelectrics**

Hasan Emre Baysal, Tzu-Yi Yu, Viktor Naenen, Stijn De Smedt, Defne Hiz, Bokai Zhang, Heyi Xia, Isidro Florenciano, Martin Rosenthal, Ruth Cardinaels, Francisco Molina-Lopez

## 1. Materials and Methods:

**Direct ink Writing (DIW) paste preparation.** For the DIW paste preparation, the method in our previous study[1] was modified. For the non-filtered material, a commercially available Poly(3,4-ethylenedioxythiophene) polystyrene sulfonate (PEDOT:PSS) aqueous dispersion (Clevios™ PH1000, Heraeus Electronic Materials) was mixed with the Li salt plasticizer bis(trifluoromethane)sulfonimide lithium salt (Alfa Aesar, CAS number: 90076-65-6) at the ratio indicated in Table S1 in an open Teflon container, and magnetically stirred overnight at room temperature. Later, the temperature was raised to 60°C, and the mixture was continuously stirred with a magnetic bar until reaching the target weight of 1.42 g (including Li salt) per 10 ml initial PEDOT:PSS dispersion. In the last step, the crosslinker (3-Glycidyloxypropyl)trimethoxysilane (GOPS) (Sigma Aldrich, CAS number: 2530-83-8) was added to the paste and stirred manually at room temperature before printing. GOPS was added at the end to prevent premature crosslinking impeaching DIW.

For the filtered samples, 15 ml of the PEDOT:PSS aqueous dispersion was diluted in 25 ml of dimethyl sulfoxide (DMSO) (Sigma-Aldrich, CAS number: 67-68-5) and stirred for 96 hours at 60°C in a closed container. The mixture was poured on top of a customized vacuum-assisted filtration and mixing system. The system continuously stirred the dispersion with a Teflon spatula to avoid film formation and aggregation on top of a 200 nm pore-size filter paper (NL16 Whatman Polyamide Membrane, Cytiva), while promoting the filtration of the extra solvent with dissolved excess PSS to be discarded (Movie S9). The dispersion was washed five times with 45 ml of deionized water, which was iteratively filtered out to remove the excess PSS and replace DMSO with water as the main dispersion medium for the subsequent freeze-drying step.

**UV-visible spectroscopy.** The filtered out excess PSS solution and a pure PEDOT:PSS dispersion were tested with a UV-Vis Spectrometer (UV1601, Shimadzu) to confirm the effective PSS removal during filtration. The test liquids were drop casted on quartz substrates and dried at room temperature to form test films.

**Rheology.** Rheological tests were conducted on an MCR501 stress-controlled rheometer (Anton Paar, Austria) with a Peltier temperature unit controlled at 25 °C. A roughened set of parallel plates with a diameter of 8 mm was used to avoid slip. The gap was set at 0.45 mm and a mineral oil ring was placed around the sample to minimize water evaporation. The linear regime of each sample was determined in terms of deformation with an oscillatory amplitude sweep at 10 rad/s and deformations ranging from 0.1 % to 100 %. Dynamic frequency sweeps in the linear regime were performed at 1 % strain. Rotational rate sweeps were conducted to create flow curves. The shear rate was increased stepwise from 0.01 $s^{-1}$ to 100 $s^{-1}$ at 90 s per shear rate, resulting in steady state flow. Additionally, creep and recovery tests were performed to test the elastic recovery after the application of a constant shear stress of 50 Pa for 1 h.

**Test film preparation.** For film preparation to test thermoelectric parameters and mechanical properties, the PEDOT:PSS aqueous dispersion (with Li salt and/or GOPS, see Table S1) was magnetically stirred for an hour, the homogeneous dispersion was drop casted on a petri dish and degassed in a desiccator for half an hour before letting it dry at ambient conditions. The dry films (with thickness values ranging from 30-80 μm) were cut in test coupons, sandwiched between two metal plates (to keep them flat), and annealed on a hot plate at 120°C for an hour to promote the GOPS-mediated crosslinking. Later, the metal-sandwiched test coupons were post-treated via ethanol vapor annealing and heat in a similar way as the aerogels.

**Mechanical testing.** The mechanical properties of these films were tested based on the ASTM norm D882. The samples were cut to a size of 70 mm x 5 mm (by scissor/rotary trimmer-Dahle 552). The samples' width and length were measured by an optical microscope (FHD Trend, Tagarno), and the thickness was measured by a digital micrometer. Universal tensile machine (UTM) tests were performed on an Instron 5943 system with air-actuated grips (Instron 2712-019): strain rate 0.05 %/s; 30-50 mm grips separation and gauge length. Samples with anomalous failure, particularly those breaking in or near the grip area, were excluded when determining the failure strain, but not for the determination of the Young's modulus. The Young's modulus was defined as the maximum tangent modulus within the linear deformation region to minimize errors from the pronounced toe-regions (i.e. the initial nonlinear regions) observed in the stress-strain curves of our samples. The toe-region was corrected to extract the strain-zero stress point point, and hence get an accurate strain at failure value according to the standard ASTMD882. Failure to follow this protocol might lead to underestimating the Young's modulus and overestimating the strain at failure.

To test the flexibility of the aerogels, the maximum bending strain was determined by bending 1.5 cm-long aerogel pillars with a diameter of 1.4 mm around cylinders with various radii. The maximum bending strain (tensile strain at the surface) was calculated with the ratio pillar radius to the real bending radius at the neutral axis. The change in resistance was recorded with the two-wires technique while bending.

**Scanning Electron Microscopy (SEM).** The morphology of the samples was characterized by an FEI XL 30 FEG scanning electron microscope. For a fair comparison, all the samples were observed at 10 kV and with no coating.

**Wide angle X-ray scattering (WAXS).** WAXS measurements were performed at the beamline BM26 (DUBBLE), European Synchrotron Radiation Facility (ESRF). Samples were probed by 12 keV X-ray with a 30-second exposure time. The sample-to-detector distance was 348.85 mm. Calibration was done with α-$Al_2O_3$ standard. A helium fly-tube was placed between the sample and detector to reduce air scattering. 2D WAXS images were collected by a Pilatus 1M detector and visualized by GIuSAXS, a visualization tool funded by DUBBLE@ESRF. WAXS data were further reduced to 1D line-cut (integrated over an azimuthal angle χ = 180°) by "BUBBLE", a data reduction software supported by DUBBLE@ESRF.

**Raman Spectroscopy.** The presented Raman spectra were collected using a Horiba LabRAM HR Evolution. Raman shift was calibrated with a NIST-certified Horiba SP-RCO calibration objective. The samples were probed with a 1.7 W, 633 nm wavelength HeNe laser with a 30-second exposure time. For each sample, the presented spectrum is the average of six spectra collected from three different spots. The spectra are fitted in Matlab by the nonlinear least-square solver *lsqcurvefit*. The pseudo-Voigt function was applied to identify peak position, intensity, and full-width-at-half-maximum (FWHM).

**Thermoelectric Characterization**

Conductivity and Seebeck coefficient, manual setup (room temperature): Electrical resistance measurements were done with the four-wires technique at room temperature with a multimeter (2000 Multimeter, KEITHLEY) connected to a probe station (SimplePS TRIAX, KeyFactor). Contact pads made of silver paste (LOCTITE® ECI 1501 E&C, HENKEL) were deposited on the aerogel sample to improve the reliability of the electrical contact. To calculate the electrical conductivity of the printed aerogel pillars, their cross-sections, and distances between the contact pads were measured with an optical microscope (Prestige Digital Microscope, Tagarno). The Seebeck coefficient was calculated with a house-made Seebeck setup. The setup consists of two Peltier elements to apply a symmetric difference in temperature (around room temperature) across the sample, thermocouples to measure the temperature, and a multimeter to measure the thermo-voltage generated through the sample. The same leads were used to measure temperature and voltage to avoid the "cold finger" effect[2]. A correction factor of + 7 µV/K was applied to the measured sample's Seebeck to account for the Seebeck coefficient effect of the leads. Contrary to the aerogel samples, no silver paste was applied to the films as it did not attach well to their surfaces.

Conductivity and Seebeck coefficient, automatic setup (temperature sweep): The room temperature values of electrical conductivity and Seebeck coefficients measured with the manual setup were confirmed with a commercial setup LSR (LSR-3, Linseis). The measurements were also extended to a wide temperature range between -10 and 90 °C. These measurements took place under a 0.05 bar He atmosphere. Measuring in an inert atmosphere ensured the absence of potential local redox reactions at the metal contact/PEDOT:PSS interface that have been recently reported to influence the Seebeck measurements when the samples are measured in the presence of humidity[3].

Thermal properties: Thermal diffusivity was measured under room temperature conditions and a nitrogen purge gas pressure of 0.5 bar with a light flash analyzer (LFA 467 Hyper Flash, Netzsch). The samples, prepared by blade coating the DIW paste followed by freeze drying, were mounted in a through-plane sample holder and coated locally with graphite spray to maximize heat absorption upon light exposure. The material's specific heat (around room temperature) was measured by differential scanning calorimetry (DSC Q2000, TA Instruments) using hermetic crucibles to prevent moisture loss, and match in this way the real-application conditions. The material density was calculated by weighting a regular-shaped sample (with well-defined dimensions and volume) on a precision scale and dividing the mass by the volume. The thermal conductivity, κ, is calculated as κ = $α·ρ·C_p$, with $α$, $ρ$, $C_P$ being the thermal diffusivity, bulk density, and specific heat, respectively (Table S2).

**Aerogel printing on a stretchable substrate.** The prepared PEDOT:PSS based hydrogel was printed with an in-house modified DIW 3D printer(modification inspired by Krige et al.[4]) on a silicone-based elastomer substrate (Ecoflex™ 00-30, Smooth-on) with an average extrusion speed of 0.5 mm/s and feed rate of 30 mm/min. To increase the adhesion between the substrate and the printed geometries, a semi-cured (mixed with crosslinker but not heat treated) Ecoflex thin film was coated on top of the fully cured substrate. Afterwards, a PLA frame was placed on the substrate and around the printed features to avoid wrinkling during the subsequent liquid nitrogen dipping and freeze-drying (in a freeze-drier Alpha 2-4 LCSbasic, Martin Christ) steps. To complete the curing of the Ecoflex substrate and the crosslinking of GOPS in the aerogel, after freeze-drying the 3D printed device was kept on a hot plate at 120°C for an hour. The devices' solvent annealing post-treatment was carried out by keeping them next to an open vial with ethanol in a desiccator for 6 hours under vacuum conditions, followed by heat treatment on a hot plate at 120°C for an hour in an ambient atmosphere.

For the stretchable conducting lines test and the planar thermoelectric demos, the 3D printed arches were interconnected/coupled with stretchable conductive silver ink (DM-SIP-2006, DYCOTEC). The silver ink was dried at room temperature overnight followed by hot-plate annealing at 60°C for 30 min in ambient conditions. In some stretchable demos, a surface mounted device (SMD) LED was connected to the circuit. The stretchability of the demos was tested with an in-house-made stretching setup consisting of two clamps which separation distance was controlled by rotating a screw. For the thermoelectric characterization of the planar thermoelectric demo, a small Peltier module was assembled in one of the clamps to heat up one side of the device.

**Dense vs. aerogel pillar's output power comparison.** First, a reference DIW paste consisting of a PEDOT:PSS commercial dispersion with 5 %vol of DMSO (labeled as sample 0*) was prepared as mentioned above (non-filtered case without additives). The 0* paste was printed and dried at ambient conditions (dense pillar). The paste 0f was freeze-dried after printing. In order to achieve the same dry pillar diameter for both pastes, the 0* paste had to be printed with a larger nozzle diameter than the 0f paste, to compensate for the substantial shrinkage suffered by 0* during ambient drying. Then, both dry pillars were post-treated by ethanol solvent annealing. For the test setup, a thermally conductive and electrically insulating substrate (RS PRO Thermally Conductive Gap Filler Material, thermal conductivity = 1.6 W/(m K)) was placed on a metal piece sitting on a Peltier unit. Silver paste (DM-SIP-2006, DYCOTEC) was used to attach both pillars to the substrate and to connect silver wires from the top and bottom ends of the pillars for electrical measurements. Different temperatures were applied to the substrate with the Peltier unit. The temperature gradient across the pillars was recorded with a thermal camera (PI640, Optris). The external temperatures applied to the pillar, i.e. the temperature of the substrate and the air above the pillar, were measured with a thermocouple attached with thermal paste on the substrate, and a thermometer (Sensirion), respectively. The output power was recorded with a source meter unit (SMU, Keysight B2901B).

**Data Handling.** For measurements with associated error bars, we reported the average of at least three data points and determined the error using the standard error (standard deviation divided by the square root of the number of measurements). For electrical conductivity and Seebeck coefficient measurements of films and aerogels, and mechanical tests of films, we sampled three specimens (or more in multiple cases) from each composition, while each Seebeck coefficient measurement involved recording the thermovoltage at three temperatures (or more in multiple cases) and performing a linear regression (the Seebeck coefficient = the fitted slope). The statistical significance of the effect that Li Salt, GOPS and post-treatment had on the electrical conductivity and Seebeck coefficient of the films was investigated with JMP software. The multiple regression models were obtained by backward subtraction using F-testing with 5% significance, the results are shown in Table S3 and Figure S4-iii.a In Figure 3c, we took as the maximum bending strain, the maximum achieved (best) for all the samples belonging to the same population (three tested samples), to account for potential structural imperfections and to test the material's maximum capability. Data in Figures 5 – 7 correspond to measurements performed on a single demo device. Errors reported for the power factor and ZT were determined from the errors of the Seebeck coefficient, electrical conductivity, and thermal conductivity using the theory of error propagation.

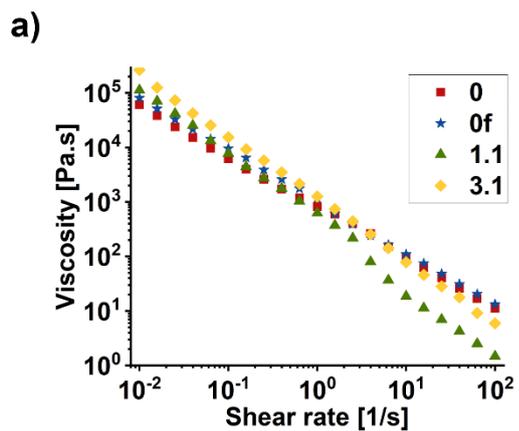
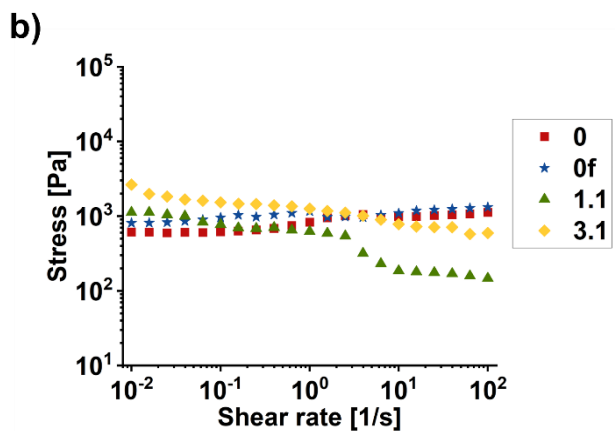
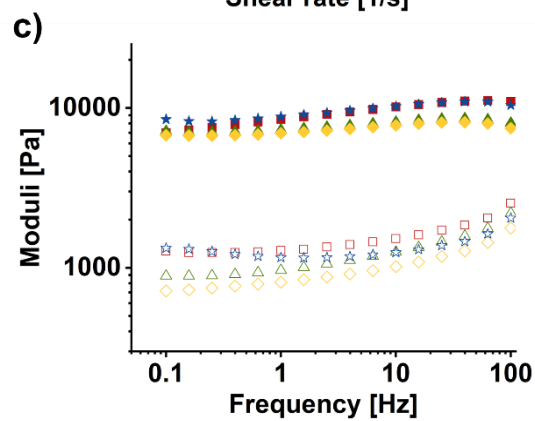
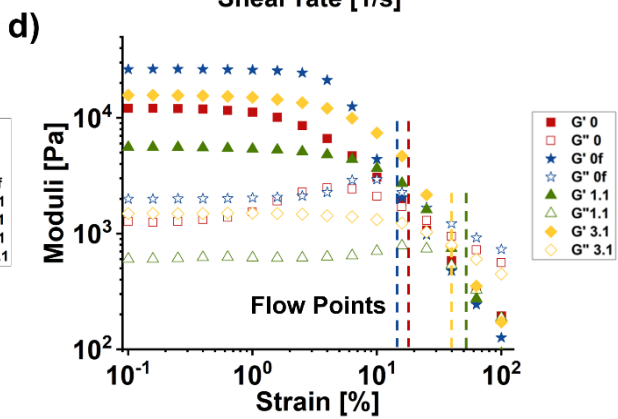
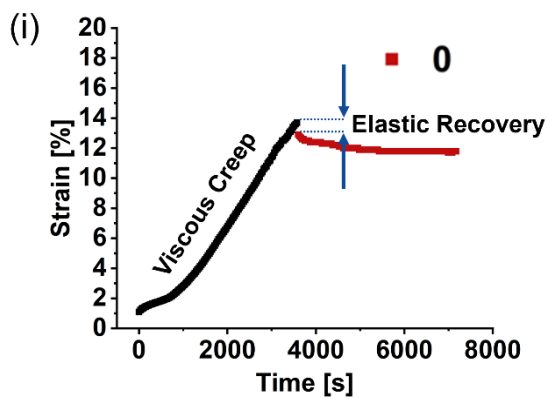
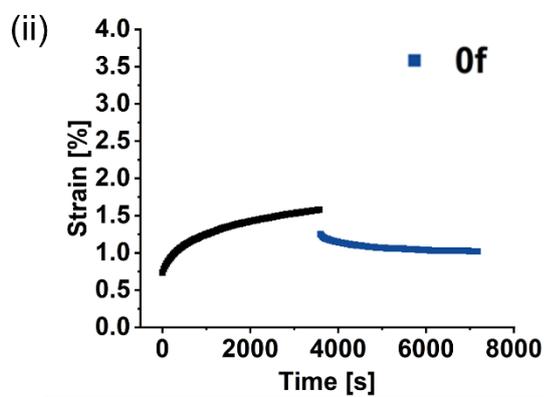
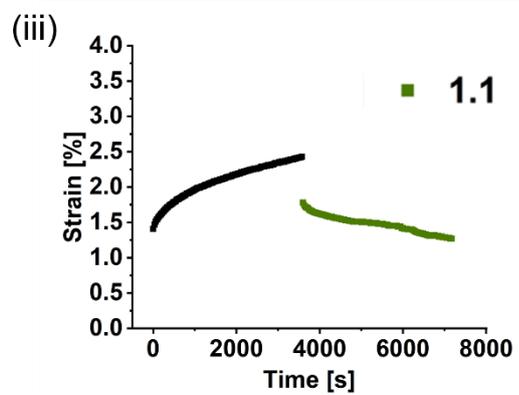
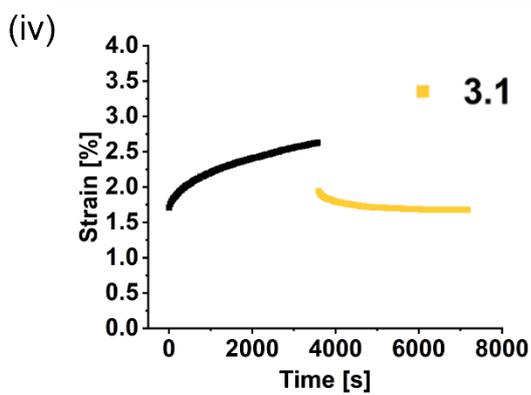

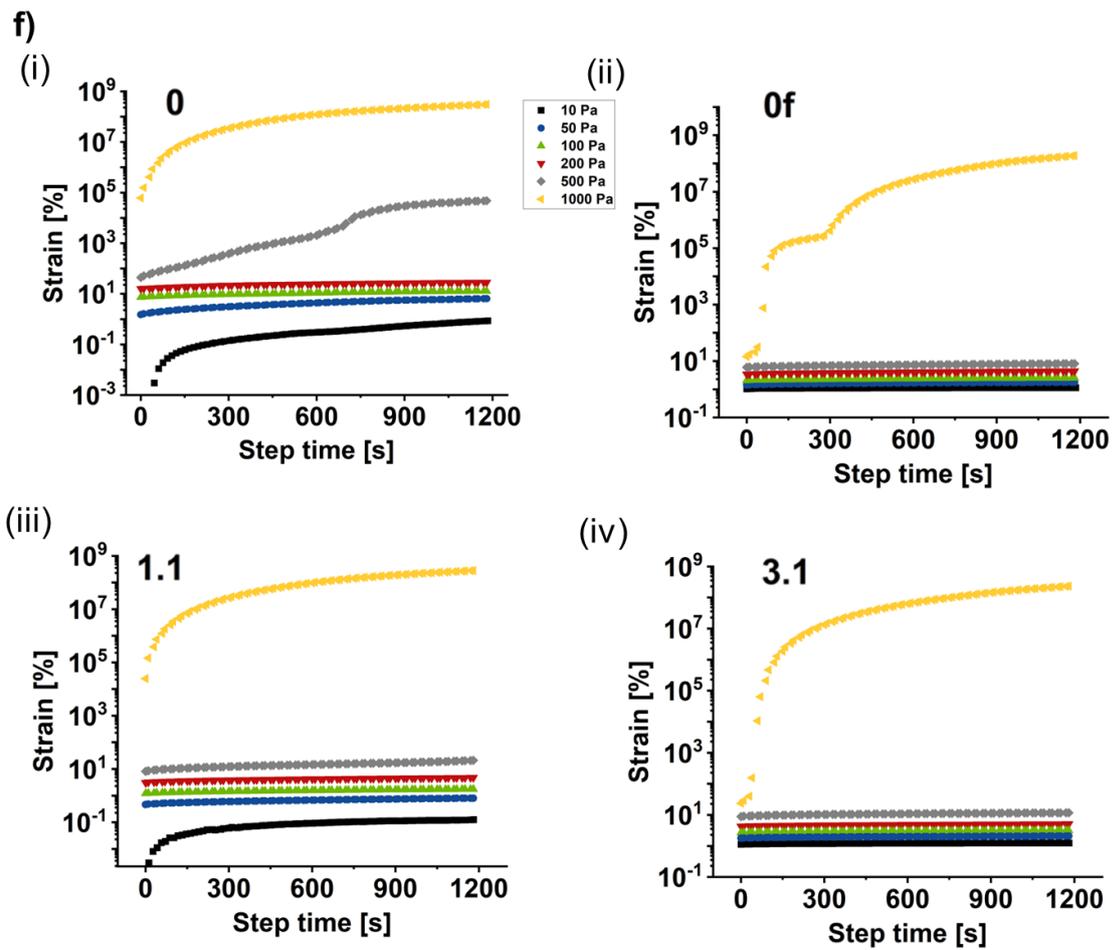

**Figure S1. Rheological characterization of selected DIW printable materials.** a) Flow curve. b) Stress change with respect to the applied shear rate. c) Frequency sweep at 1% strain. d) Strain sweep at 10 rad/s. e) Creep-recovery test. f) Stress sweep.

## 2. Rheology considerations

To determine the relevant shear rate range for the flow sweeps (Figure S1a), an order of magnitude estimate of the shear rate at the nozzle wall in the 3D printing process can be obtained from the printing flow rate. For estimation purposes, we assume a constant viscosity and absence of wall slip, which results in a parabolic flow profile in the nozzle with zero velocity at the wall. The flow rate follows from the extrusion velocity, *v*, and plunger surface area, *A*, as:

$$Q = v \cdot A \approx 10^{-8} \ m^3/s \qquad (Eq. S1)$$

Subsequently, the maximum shear rate, which is the shear rate at the wall, can be obtained as [1]:

$$\dot{\gamma} = \frac{4Q}{\pi R^3} \approx 30 \ 1/s \qquad (Eq. S2)$$

We assume that the DIW paste is a Herschel-Bulkley fluid, with shear stress $\tau$ given by:

$$\tau = \tau_0 + k \dot{\gamma}^n \qquad (Eq. S3)$$

This results in the following expression for the viscosity $\eta$:

$$\eta = \frac{\tau}{\dot{\gamma}} = \frac{\tau_0}{\dot{\gamma}} + k \dot{\gamma}^{n-1} \qquad (Eq. S4)$$

where $\tau_0$ denotes the yield stress, k stands for the consistency index, $\dot{\gamma}$ represents the shear rate, and *n* denotes the flow index. When $\tau$ is smaller than $\tau_0$, the Herschel-Bulkley fluid exhibits the characteristics of a rigid (non-deformable) solid. Otherwise, it behaves as a fluid. For *n* values less than 1, the fluid demonstrates shear thinning behavior, while for n values greater than 1, the fluid exhibits shear thickening behavior. When n equals 1 and $\tau_0$ equals 0, this model simplifies to that of a Newtonian fluid. As can be observed from Figures S1a and S1b, within the accessible range of shear rates the slope of the viscosity versus shear rate curve (on log-log scale) is -1, which means that the flow curve is dominated by the yield stress $\tau_0$, which is around 1000 Pa for all samples (Fig. S1b).

The high yield stress of the hydrogel allows it to withstand deformations of the printed structure due to gravity and surface tension. The surface tension effects, driving retraction of printed filaments, can be quantified by the Laplace pressure [2]:

$$\Delta P = \frac{2\gamma}{R} \qquad (Eq. S5)$$

with the surface tension γ ~ 70 mN/m (aqueous paste) and the nozzle radius R ~ 800µm, this results in a Laplace pressure ΔP of ~175 Pa. Thus, the yield stress is largely sufficient to avoid retraction due to the surface tension. The effect of gravity in printability can be determined by the following condition: [3]:

$$1 > \frac{\rho g h}{2\tau_0} \qquad (Eq. S6)$$

Where ρ is the density of the material, *g* is the gravity acceleration and *h* is the height of a virtual column of liquid on top of the material. This dimensionless ratio, which resembles the plastic Galilei number, represents the ratio of the gravitational stress to the yield stress. According to Eq. S6 it is expected that the bottom layers of our aerogel pillars can carry up to a column 0.2 m height of printed material on top before it yielding.

Figures S1c and S1d demonstrate the elastic behavior of all hydrogels at low strains, with a transition to more viscous behavior at strains above 10%.

Figure S1e shows the creep-recovery response of the hydrogels, whereby a stress of 50 Pa is applied for 3600 s, followed by recovery under zero stress. During the application of the stress, it can be seen that a large portion of the deformation originates from viscous creep, even below the yield stress/strain. All materials exhibit a similar amount of limited immediate elastic recovery of ~1% when the applied stress is released (downwards jump in strain in Figures S1e (i)-(iv) at 3600s). The addition of Li salt (sample 1.1 and 3.1) and the freeze drying (sample 0) make the material less deformable under the application of a constant stress below the yield stress while not altering other rheological properties significantly (comparison of the maximal strain in Figures S1e (i) with (ii)-(iv) and rheological results in S1e a, b and c). Figure S1f shows stress sweep tests, with each stress being applied for 1200s. These confirm that yielding occurs between 500 and 1000 Pa, with sample 0 having a slightly lower yield stress value and more viscous creep at low stress values.

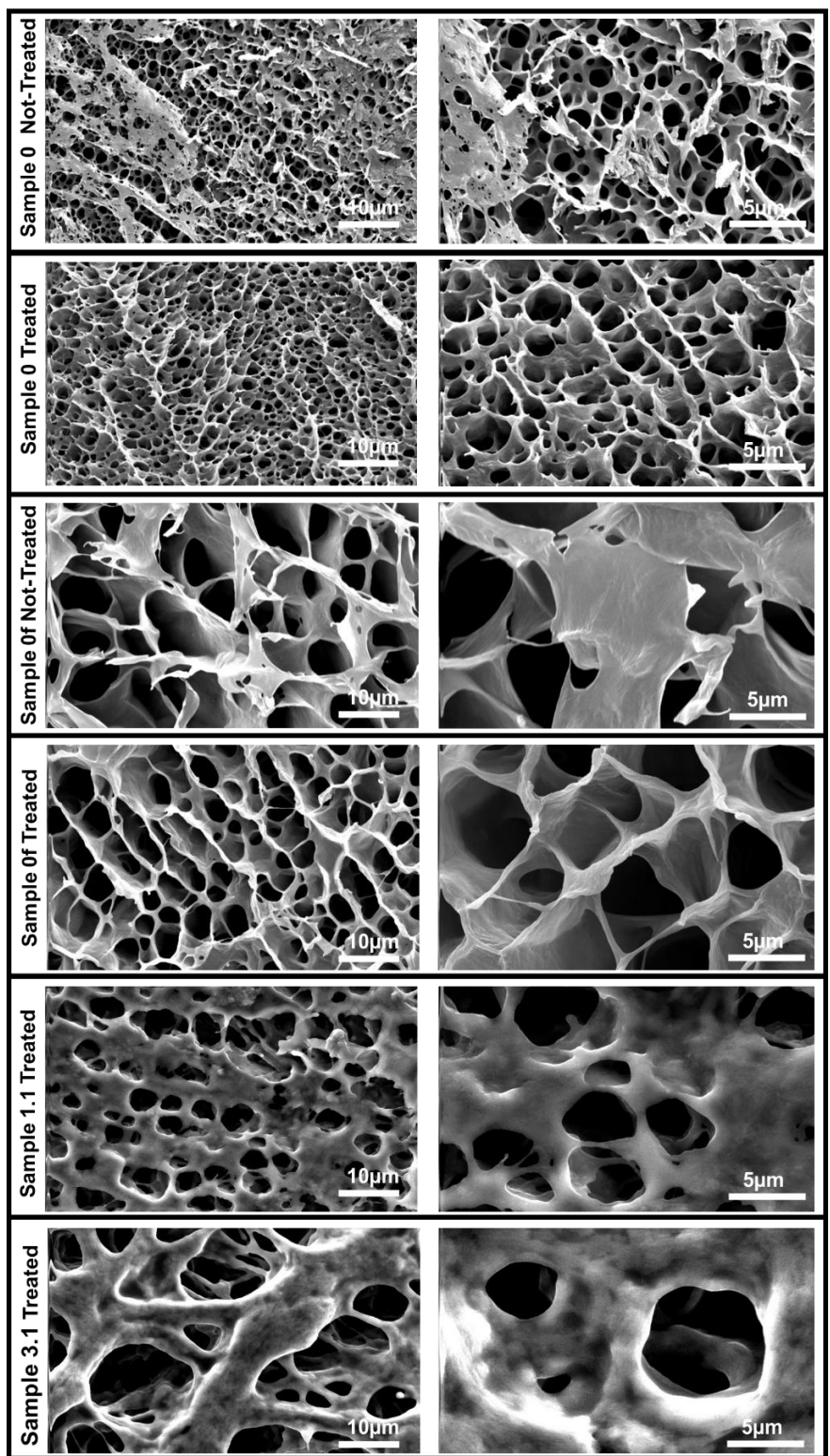

**Figure S2. Scanning Electron Microscopy (SEM) images of the printed aerogels with selected compositions.** The SEM images include both materials before and after ethanol vapor annealing post-treatment. The images reveal that the post-treatment respected the original porosity of the material and improved its surface by welding loose edges (presumably rich on PSS).

**Table S1. Material composition and corresponding labeling.** According to the manufacturer of the PEDOT:PSS dispersion (Heraeus Clevios PH1000), the solid content of the dispersion is 1-1.3 wt%.

| Label | PEDOT:PSS (ml) | DMSO (ml) | GOPS (mg) | Li Salt (mg) |
|---|---|---|---|---|
| 0 | 10 | - | - | - |
| 0* | 10 | 0.2 | - | - |
| 0f | 10 (filtered excess PSS) | - | - | - |
| 0.1 | 10 | - | - | 33 ± 3 |
| 0.2 | 10 | - | - | 66 ± 3 |
| 0.3 | 10 | - | - | 140 ± 5 |
| 0.4 | 10 | - | - | 200 ± 5 |
| 1 | 10 | - | 6 ± 1 | - |
| 1.1 | 10 | - | 7 ± 1 | 142 ± 5 |
| 2 | 10 | - | 18 ± 1 | - |
| 2.1 | 10 | - | 17 ± 1 | 33 ± 3 |
| 2.2 | 10 | - | 17 ± 1 | 200 ± 5 |
| 3 | 10 | - | 33 ± 3 | - |
| 3.1 | 10 | - | 28 ± 3 | 140 ± 5 |
| 3.2 | 10 | - | 33 ± 3 | 209 ± 5 |

**Table S2. Parameters used for thermal conductivity calculation.**

| Aerogel | Density [g/cm$^3$] | Diffusivity [mm$^2$/s] | Cp [J/(g K)] | Conductivity [W/(m K)] |
|---|---|---|---|---|
| 0 | 0.103 ± 0.002 | 0.746 ± 0.002 | 1.80 | 0.1397 ± 0.0013 |
| 0f | 0.110 ± 0.002 | 0.358 ± 0.005 | 1.65 | 0.0651 ± 0.0008 |
| 1.1 | 0.280 ± 0.004 | 0.328 ± 0.003 | 1.84 | 0.1695 ± 0.0016 |
| 3.1 | 0.231 ± 0.002 | 0.286 ± 0.002 | 1.79 | 0.1186 ± 0.0007 |

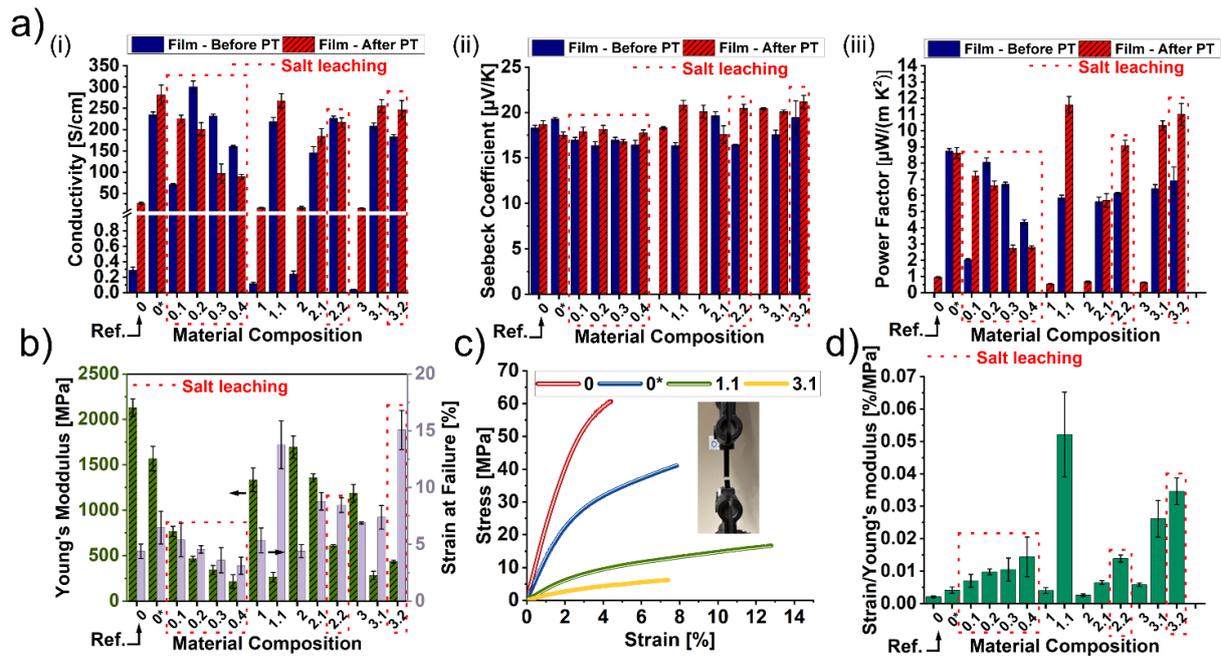

**Figure S3. Thermoelectrical and mechanical characterization of thick films for material selection (before and after post-treatment, measurements performed at ambient conditions).** a) Electrical conductivity (i), Seebeck coefficient (ii), and power factor (iii). b) Young's modulus and strain at failure of post-treated films. c) Representative stress-strain curves of post-treated materials. d) Ratio of strain at failure to Young's modulus for post-treated films. For detail about material composition refer to Table S1.

## 3. Significant effect of the additives

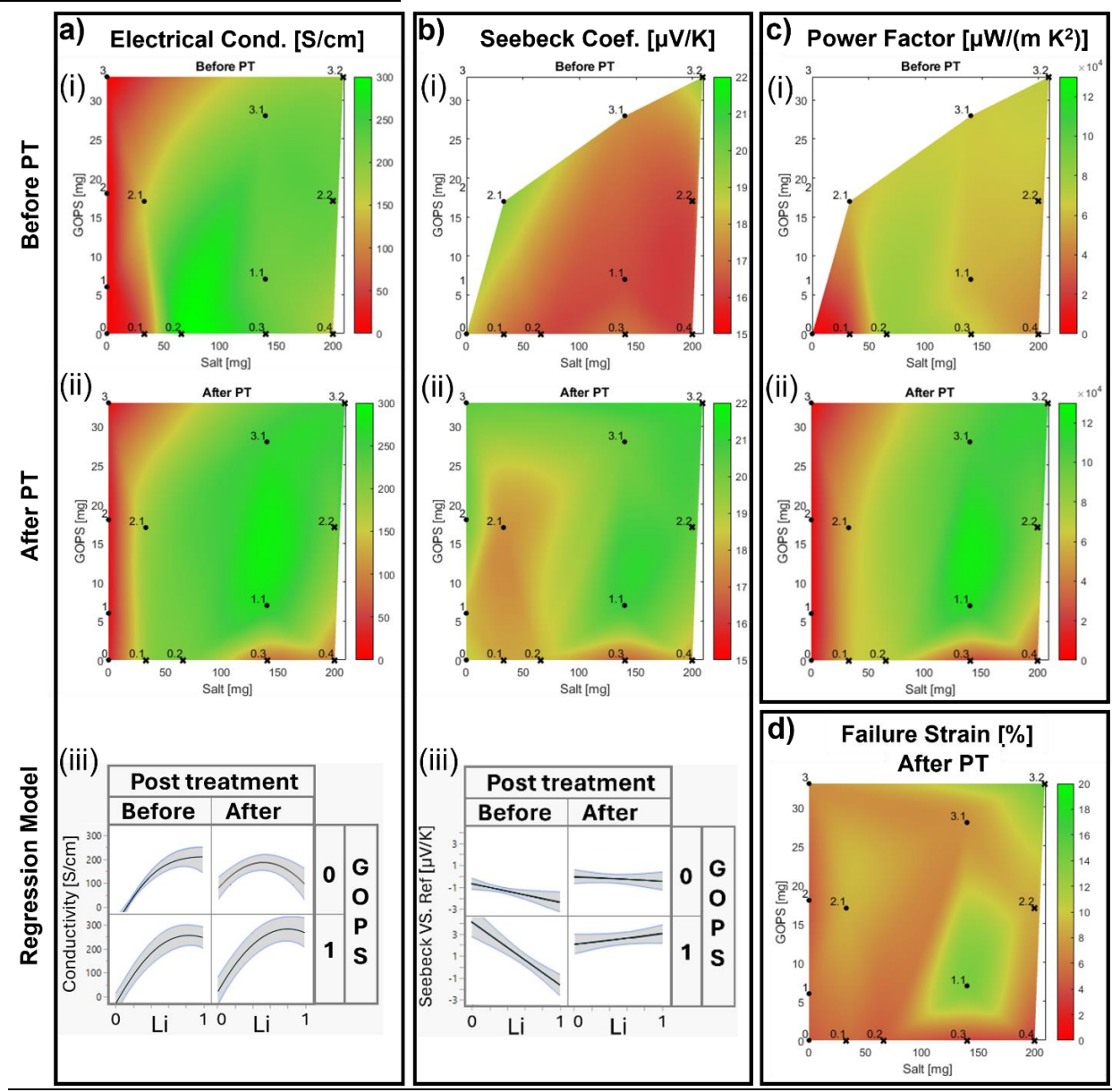

**Figure S4. Thick film thermoelectric and mechanical performance color plots and statistical analysis for varying Li salt and GOPS concentrations.** a) Color plot depicting conductivity before (i) and after (ii) post-treatment (PT) and the regression model profiles (iii). b) Color plot depicting Seebeck coefficient before (i) and after (ii) post-treatment and the regression model profiles (iii). c) Color plot depicting power factor before (i) and after (ii) post-treatment (PT). d) Color plot depicting failure strain after post-treatment (PT). The color plot utilizes cubic interpolation to smooth the discrete data from Figure S3's data. Each formulation is marked in the design space by the marker "•" (or "x" in the case of Li salt leaching). In the regression model profile plots, the maximum and minimum amount of Li salt and GOPS are indicated by 1 and 0, respectively.

To evaluate the statistical significance of the effects of Li salt, GOPS, and post-treatment on the electrical conductivity and Seebeck coefficient of thick films, a multivariable regression based on backward subtraction (F-test, significance level of 5%) was employed. The variables "amount of Li salt" and "amount of GOPS" were coded from 0 to 1 so that no interaction is possible when one of the additives is absent, to accurately reflect physical conditions. The pristine PEDOT:PSS film (formulation 0) without post-treatment with all variables at the low level served as the reference and was subtracted from all others, effectively excluding the intercept term from the analysis.

**Conductivity (σ):** The response surface in Figure S4a and the regression model in Table S3 indicates that there is a strong and positive influence of Li salt on conductivity (coefficient = 326), reaching an maximum value before declining due to the detrimental effect of salt leaching at high salt mass ratio. Note that leaching leads physically to an actual decrease in the amount of salt within the material. This quadratic trend of Li salt mass ratio and conductivity is underpinned by the statistically significant quadratic Li*Li effect ($P < 0.0001$) in Table S3. While GOPS exhibited a statistically significant effect ($P = 0.035$), it was comparatively smaller (coefficient = 50) than Li and Li*Li. Post-treatment (PT) exhibited also a significant and overall slightly positive effect on conductivity (coefficient = 38, $P = 0.0106$). Even though there was a clear negative influence on conductivity of combining post-treatment with the use of Li salt (PT*Li, coefficient = -172, $P < 0.0001$), the further addition of GOPS led to a clear positive synergy (PT*Li*GOPS, coefficient = 227, $P = 0.0029$). This synergy is evident in Figure S4a, where post-treatment induced a shift in the optimal region of conductivity towards higher Li salt and GOPS amounts.

In summary, the regression analysis and color plot indicate that while Li salt greatly benefits conductivity, it is susceptible to leaching, particularly in large concentrations and during post-treatment. GOPS alone offers a minor direct benefit, but synergizes with Li salt by stabilizing it, leading to optimal performance at elevated levels of both Li salt and GOPS. This synergy requires ethanol vapor post-treatment.

**Seebeck coefficient - *S*:** Figure S4b and Table S3 illustrate the limited, albeit statistically significant ($P < 1\%$ in all cases), effect of Li salt, GOPS and PT on the Seebeck coefficient. Post-treatment is found to be crucial to turn the adverse effects of Li salt and the interaction Li*GOPS (coefficients = -3.1 and -4.2, respectively), into a positive three-way PT*Li*GOPS synergy (coefficient = 5.5).

**Power factor:** Consistent with the previous analysis, Figure S4c and Table S3 illustrate that the highest power factor (calculated as $S^2 \cdot$conductivity) is achieved in films that have undergone post-treatment and contain high levels of Li salt and GOPS, and did not display salt leaching.

**Mechanical – Failure Strain (only for films after PT):** The color plot (Figure S4d) and regression analysis (Table S3) for failure strain of films after PT did not reveal significant correlations or second-order interactions, only a main positive effect from both Li salt and GOPS.

**Table S3. Statistical analysis of the effects of additives and post-treatment on the material thermoelectric and mechanical performance.** The study includes the effect of Li Salt (Li), GOPS, post-treatment (PT), and their combinations on electrical conductivity, Seebeck coefficient, and failure strain. (DoF: degree of freedom).

| Electrical Conductivity | | | | Seebeck Coefficient | | | | Failure Strain | | | |
|---|---|---|---|---|---|---|---|---|---|---|---|
| Effect (DoF = 68) | F Ratio | Coef. | P | Effect (DoF = 58) | F Ratio | Coeff. | P | Effect (DoF = 68) | F Ratio | Coeff. | P |
| Li | 151.3 | 326 | <0.0001 | Li | 85.2 | -3.1 | <0.0001 | Li | 10.4 | 3.9 | 0.0020 |
| Li*Li | 23.5 | - 409 | <0.0001 | GOPS | 64.1 | 2.8 | <0.0001 | GOPS | 4.2 | 2.6 | 0.0448 |
| GOPS | 4.6 | 50 | 0.0350 | Li*GOPS | 7.2 | -4.2 | 0.0095 | | | | |
| PT | 6.9 | 38 | 0.0106 | PT | 30.9 | 1.2 | <0.0001 | | | | |
| PT*Li | 20.8 | - 172 | <0.0001 | PT*Li | 37.0 | 3.2 | <0.0001 | | | | |
| PT*Li*GOPS | 9.5 | 227 | 0.0029 | PT*Li*GOPS | 8.8 | 5.5 | 0.0043 | | | | |

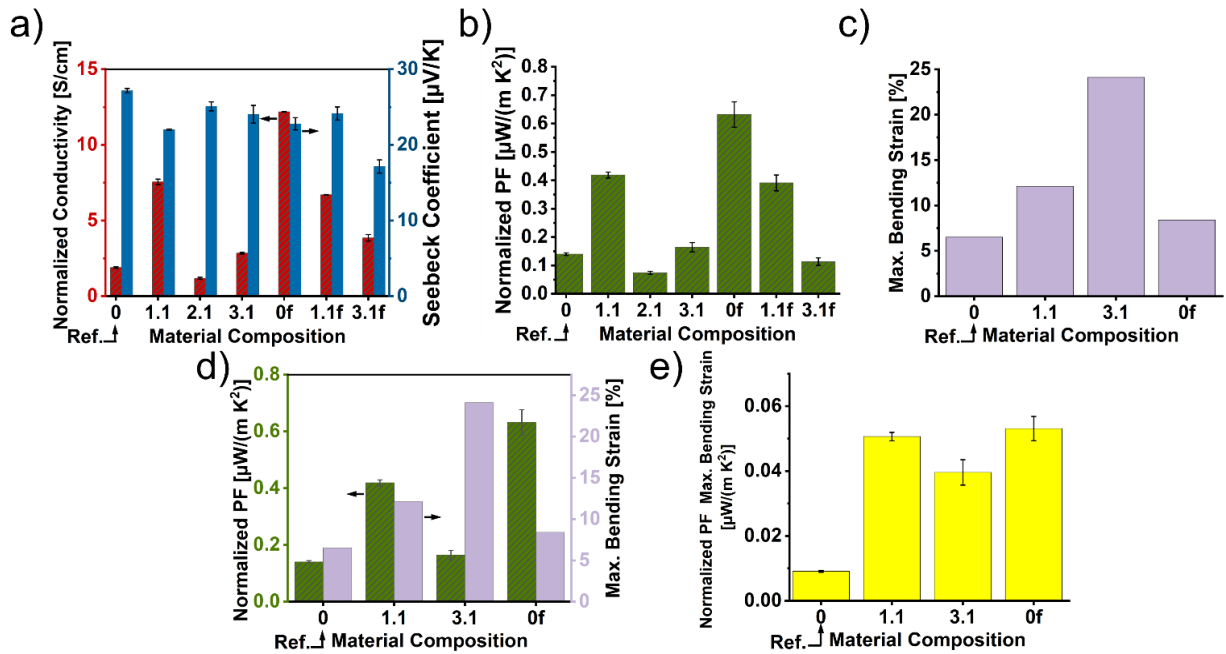

**Figure S5. Normalized (to density corresponding to 90%vol of air) thermoelectrical and mechanical characterization of 3D printed aerogel filaments for material selection (after post-treatment, measurements performed at ambient conditions).** a) Normalized electrical conductivity and Seebeck coefficient. b) Normalized power factor. c) Maximum bending strain to which the parts survive upon bending. d) Normalized power factor and maximum bending strain. e) New electromechanical figure of merit: ratio of normalized power factor to the maximum bending strain.

# 4. Explanation of the conductivity normalization to the aerogel expected density

The density normalization was employed to adjust for the varying volume fraction of air in each aerogel sample, which arose from the a poor control in the total water content in the paste before the freeze-drying step. As a result, the final material density varied significantly for each material composition and slightly for different batches of the same composition (Table S2). To mitigate this variability and enable a fairer comparison among aerogels, we utilized the deviation of each material from its expected density to compensate for conductivity. The expected density was calculated assuming that 90% of the aerogel volume was air, which was roughly the case for most measured aerogels. Consequently, materials with a higher density than expected (indicating a comparatively higher ratio of dense phase) were penalized accordingly in terms of conductivity. A similar correction was reported before by other groups in a similar context[5].

The justification for this normalization is the following. We assume the aerogel to be a composite of a (percolated) conducting phase (polymer with conductivity $\sigma_{polymer}$) and insulating phase (air with conductivity $\sigma_{air} \sim 0$). Thus, the aerogel conductivity should follow a general rule of mixture[6]:

$$\sigma^{\alpha}_{aerogel} = f\sigma^{\alpha}_{PEDOT:PSS} + (1-f)\sigma^{\alpha}_{air} \approx f\sigma^{\alpha}_{PEDOT:PSS} \qquad (Eq.\,S7)$$

where $f$ is the volume fraction of the polymer and $-1<\alpha<1$ is a fitting parameter that depends on the composite morphology. Assuming $\alpha =1$, which is true only for polymer fibers perfectly parallel to the electric field, Eq. S7 stablishes that the aerogel conductivity is proportional to the polymer volume fraction. Likewise, the aerogel density ($\rho_{aerogel}$) is also proportional to the polymer volume fraction: $\rho_{aerogel} \sim f\rho_{polymer}$. Hence $\sigma_{aerogel} \propto \rho_{aerogel}$.

Although this normalization involves a strong assumption in the aerogel morphology (all polymer fibers parallel to transport) and is likely introducing some degree of error, it resulted useful for a rough qualitative analysis. Indeed, after density correction, a similar trend to that observed for films in Figure 2 was noted for the aerogels in Figure S5. Note that density normalization was done exclusively for material screening and no claims of absolute performance were made based on normalized values. The results shown in Figure 3 correspond to actual (non-normalized) measurements.

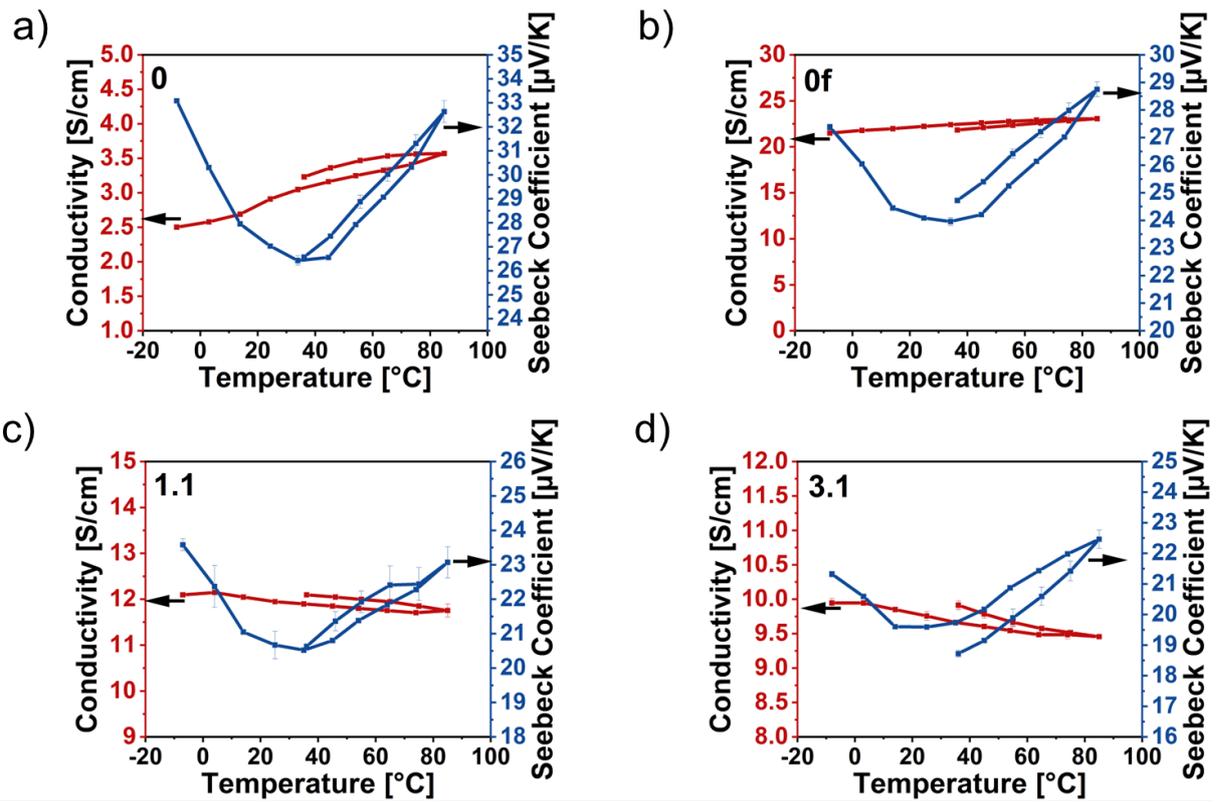

**Figure S6. Electrical conductivity and Seebeck coefficient change with temperature (between -10 °C and 90 °C) for different PEDOT:PSS aerogels.** a) Sample 0. b) Sample 0f. c) Sample 1.1. d) Sample 3.1. The tested range of temperature is relevant for wearables and applications related to the IoT.

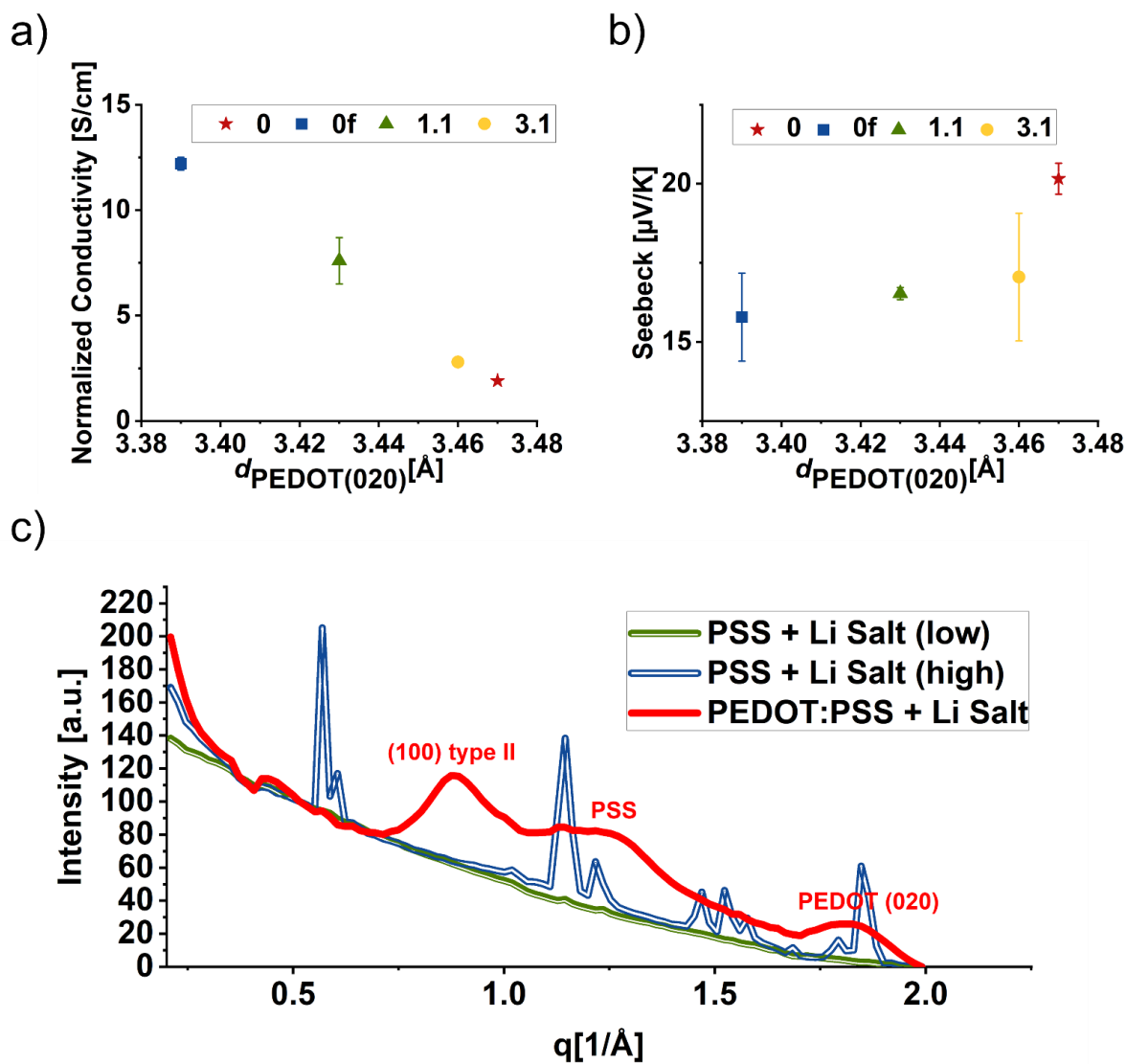

**Figure S7. Synchrotron X-ray diffraction study.** a) Electrical conductivity, and b) Seebeck coefficient versus PEDOT π–π stacking distance -peak (020)- measured with Wide-Angle X-ray Scattering (WAXS) in transmission mode. c) Grazing-incidence WAXS (GIWAXS) 1D line-cut intensity profile of PEDOT:PSS with Li salt films and bare PSS with Li salt films (taken as reference).

## 5. Thermoelectric output power calculations for the stretchable planar device (aerogel sample # 1.1)

**Table S4.** *Device and material parameters for the stretchable planar device.* The values of conductivity and Seebeck for the aerogel 1.1. are extracted from Figure 3a. The values of conductivity and Seebeck for the stretchable silver were taken from the manufacturer and measured in our lab, respectively.

| Length of the aerogel leg: | L = 6.7 cm |
|---|---|
| Aerogel filament cross sectional area (circular): | A = 1.5 10$^{-2}$ cm$^2$ (radius = 0.14cm) |
| Electrical conductivity of Aerogel 1.1 (p-type leg): | $\sigma_{aerogel\ 1.1}$ = 16 S/cm |
| Seebeck coefficient of Aerogel 1.1 (p-type leg): | $S_p$ = 23.5 µV/K |
| Electrical conductivity of Silver line (n-type leg): | $\sigma_{silver}$ = 2 10$^4$ S/cm |
| Seebeck coefficient of Silver line (n-type leg*): | $S_n$ = 6.5 µV/K |

*Even though the silver was used in the n-type leg, it is actually a weak p-type material.

The maximum output power ($P_{max}$) provided by of a thermocouple harvester is[7,8]:

$$P_{max} = \frac{V_{OC}^2}{4R_i} = \frac{[(S_p - S_n)\Delta T_L]^2}{4(R_p + R_n + R_c)} \qquad (Eq.\,S8)$$

Where $V_{oc}$ is open circuit voltage; $S_p$ and $S_n$ are the Seebeck coefficient of the used p-type and n-type materials, respectively; $\Delta T_L$ is the externally-applied temperature difference across the device length; and $R_i$, $R_p$, $R_n$, $R_c$ are the following electrical resistances: total internal resistance of the thermocouple, resistance of the p-type leg, resistance of the n-type leg, and total contact resistance. Hence, $R_i = R_p + R_n + R_c$.

The total internal resistance of the device, $R_i = R_p + R_n + R_c$ = 27.5 Ω, can be extracted from the slope of the measured I-V curve presented in Figure 6b. The resistance of the PEDOT:PSS aerogel p-type leg is, according to the Pouillet's law:

$$R_p = \frac{1}{\sigma}\frac{L}{A} = 27.3\ \Omega \qquad (Eq.\,S9)$$

(Refer to Table S4 for the values of σ = $\sigma_{aerogel\ 1.1,}$ L and A)
From the calculatiosn above, it can be concluded that $R_i$ is dominated by the p-type leg, $R_p$, and the summed contribution of the stretchable Ag used as n-type leg and the contact resistances is negligible: $R_n + R_c$ = 0.2 Ω. Thus, it can be assumed for simplicity that $R_n \simeq R_c \simeq 0$. The fact that $R_n$ is negligible compared to $R_p$ is not surprising in view of the much higher electrical conductivity of the stretchable Ag paste compared to the PEDOT:PSS aerogel 1.1 ($\sigma_{silver}$ = 2 10$^4$ S/cm Vs $\sigma_{aerogel\ 1.1}$ = 16 S/cm).

Taken into account the previous considerations, Eq. S8 becomes:

$$P_{max} \approx \frac{[(S_p - S_n)\Delta T_L]^2}{4R_p} \qquad (Eq.\,S10)$$

To understand the mechanical strain invariability of the output power observed for the stretchable device in Figure 6d, a strain-dependent version of Eq. S10 is proposed:

$$P_{max}(\varepsilon) = \frac{[(S_p(\varepsilon) - S_n(\varepsilon))\Delta T_L]^2}{4\left(R_p(\varepsilon)\right)} \qquad (Eq.\,S11)$$

The Seebeck coefficient has been demonstrated to be invariant to the applied strain[7]. Hence: $S_{p,n}(\varepsilon) \simeq S_{p,n}$. The change in electrical resistance of the arched PEDOT:PSS (# 1.1) p-type leg is negligible ($\Delta R/R_0$ < 1.5%) below 10% of applied strain as shown in Figure 5b. Accordingly $R_p(\varepsilon) \simeq R_p$ and Eq. S11 becomes Eq. S10, demonstrating that the output power is, in practice, strain-independent up to ~ 10% strain as shown in the Figure 6d.

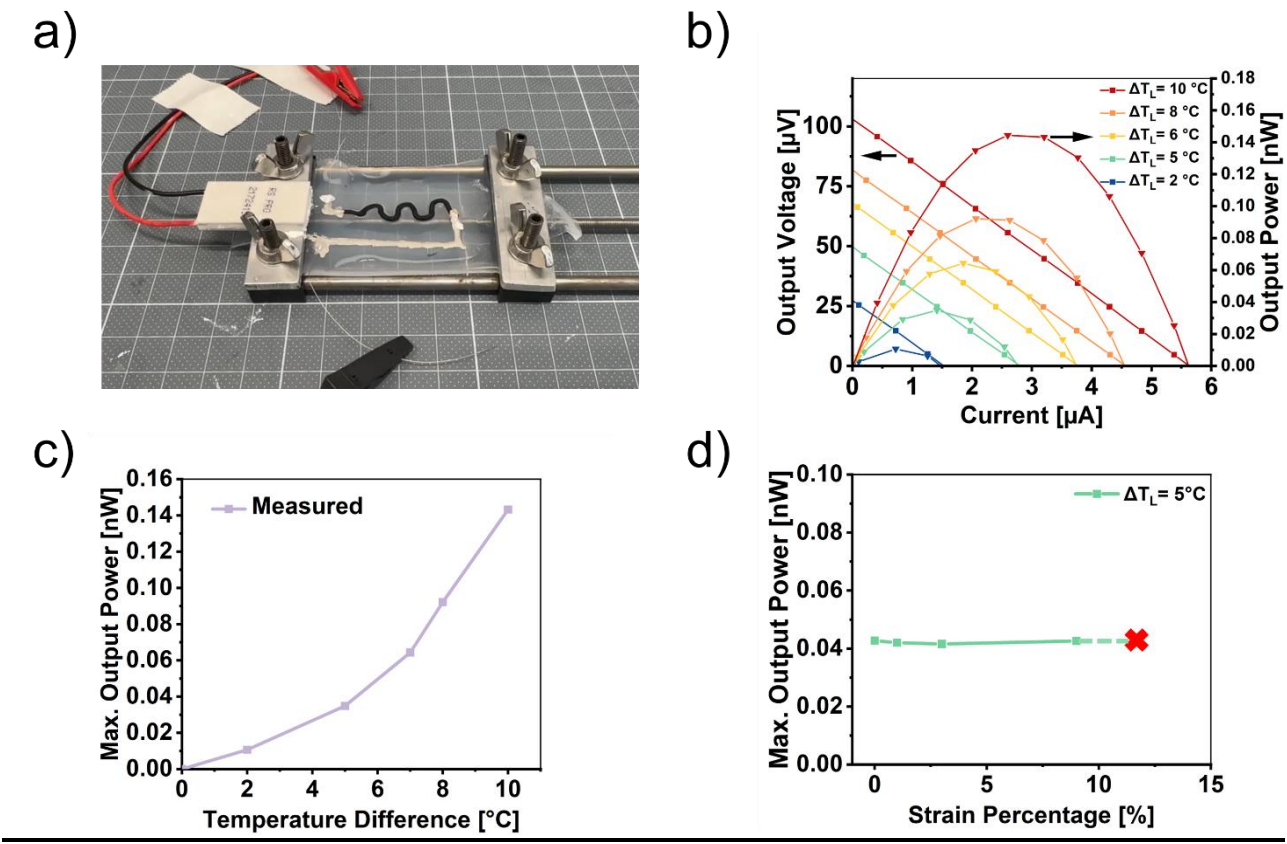

**Figure S8. Stretchable planar thermoelectric generators based on in-plane printed serpentine.** a) Optical photograph of a planar stretchable thermocouple composed of in-plane PEDOT: PSS (material composition 1.1) arches and painted stretchable silver. The device is mounted in a house-made stretching set-up and coupled to a Peltier unit at one end to generate a thermal gradient. b) Output voltage and output power vs. output current for different resistive loads and temperature differences across the device. c) Maximum power output at different temperature differences across the device. d) Output power evolution vs. applied strain for a 5°C degree temperature difference across the device.

# 6. Thermoelectric output power calculations for the vertical device (aerogel sample # 0f Vs dense sample # 0*)

To calculate the theoretical maximum output power of the dense ($P_{max,dense}$) and aerogel ($P_{max,aerogel}$) thermoelectrical pillar legs with free cold side, a single-leg variation of Eq. S8 can be drawn:

$$P_{max} = \frac{V_{OC}^2}{4R_i} = \frac{[S\Delta T_L]^2}{4(R_L + R_c)} \qquad (Eq.\,S12)$$

Where the total internal electrical resistance ($R_i$) of the pillar is the sum of the leg electrical resistance ($R_L$) and contact electrical resistance ($R_C$). This is $R_i = R_L + R_C$ according to the electrical circuit model in Figure S10a. Note that the contact electrical resistance $R_C$ is composed of the top ($R_{C1}$) and bottom ($R_{C2}$) contact resistance: $R_C = R_{C1} + R_{C1}$. Using Pouillet's law (as in Eq. S9) for $R_L$, we can express $R_i$ as:

$$R_i = \frac{1}{\sigma}\frac{L}{A} + R_C \qquad (Eq.\,S13)$$

The thermal resistance ($R'_L$) of the pillar leg can be expressed with the thermal equivalent of Eq. S9 as:

$$R'_L = \frac{1}{\kappa}\frac{L}{A} \qquad (Eq.\,S14)$$

Where κ is the thermal conductivity of the pillar material.

The total heat ($Q$) transferred through the material is, according to the Fourier's law for an isotropic medium, given by:

$$Q = \frac{\Delta T_E}{\sum_i R'_i} = \frac{\Delta T_E}{R'_H + R'_L + R'_C} \qquad (Eq.\,S15)$$

Where $R'_H$ and $R'_C$ are the thermal contact resistance at the hot (substrate-pillar interface) and the cold (leg-air interface) side, respectively (see thermal circuit model in Figure S10b). In this analysis, the Thomson effect, and the Joule and Peltier heating can be safely neglected (a "back of the envelope" calculation supports this simplification). The conduction and convection heat losses to the air have been also neglected for the sake of analytical simplicity. Therefore the same heat flows through the hot and the cold side of the pillar. It is worth noting that the assumption of negligible conduction heat to the air holds well for materials which thermal conductivity is much higher than the air's. In the case at hand, however, this simplification may introduce a certain error as $\kappa_{aerogel}$ = 0.065 W m$^{-1}$ K$^{-1}$ is comparable to $\kappa_{air}$ = 0.026 W m$^{-1}$ K$^{-1}$ (at ambient conditions)[9]. This means that caution should be exercised when performing quantitative analysis, but this limitation should take nothing away from the validity of the qualitative analysis intended in this work.

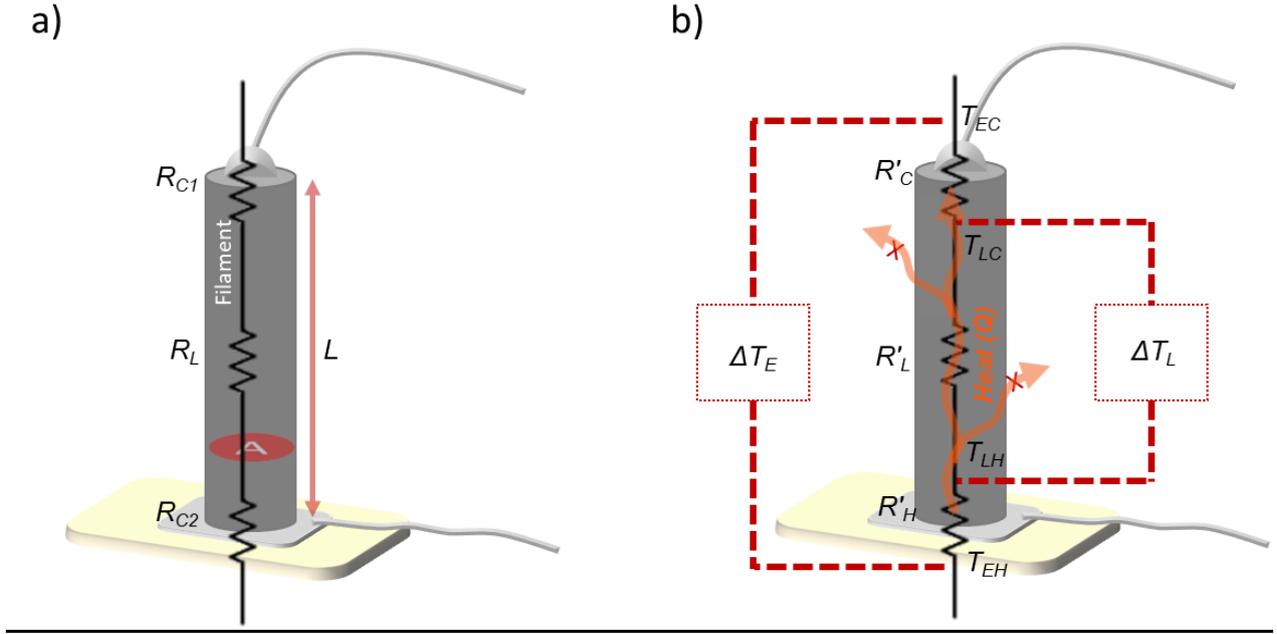

**Figure S9. Vertical pillar legs deposited on a heated substrate and with a natural air-cooled top.** a) Electrical circuit model b) Thermal circuit model.

According to Fourier's law, the drop in temperature across the pillar $\Delta T_L = R'_L Q$. Then, using Eq. S15, $\Delta T_L$ can be expressed as a function of the externally applied temperature difference **($\Delta T_E$) as:**

$$\Delta T_L = R'_L Q = \frac{R'_L}{R'_H + R'_L + R'_C} \Delta T_E \qquad (Eq.\,S16)$$

Substituting in the equation above $R'_L$ for Eq. S14 yields:

$$\Delta T_L = \frac{\frac{1}{\kappa}\frac{L}{A}}{R'_C + R'_H + \frac{1}{\kappa}\frac{L}{A}} \Delta T_E = \frac{\frac{L}{A}}{\kappa(R'_C + R'_H) + \frac{L}{A}} \Delta T_E \qquad (Eq.\,S17)$$

Finally, replacing Eq. S13 and Eq. S17 in Eq. S12, we get an expression for the maximum output power as a function of contact resistances, material properties and geometry:

$$P_{max} = \frac{(S\,\Delta T_E)^2 \left[\dfrac{\frac{L}{A}}{\frac{L}{A} + \kappa(R'_H + R'_C)}\right]^2}{4\left(\dfrac{1}{\sigma}\dfrac{L}{A} + R_C\right)} \qquad (Eq.\,S18)$$

To illustrate the effect of the contact resistances in the performance of TEG, the maximum power output at 300 K is calculated based on Eq.S18 for four different scenarios, accounting for various contact resistances: no contact resistances, only thermal contact resistance, only electrical resistance, and a combination of both thermal and electrical contact resistances. In the absence of any contact resistance, a comparison of an aerogel (like 0f) with $ZT \sim 2.4 \times 10^{-3}$ ($\sigma$ = 13 S/cm and $\kappa$ = 0.065 W/(m·K), S=20 µV/K), with a dense material (like 0*) having $ZT \sim 6 \times 10^{-3}$ ($\sigma$ = 280 S/cm and $\kappa$ = 0.55 W/(m·K), S=20 µV/K) reveals that an aerogel material would produce more than 20 times less power than a dense material: 8 vs. 181 nW (Figure S11a). However, when considering only thermal (Figure S11b) or electrical contact resistance (Figure S11c), the power output difference decreases to less than 3 times or only 1.1 times, respectively. Remarkably, when both contact resistances are taken into account, the aerogel's power output surpasses that of the dense structure by 7 times (0.42 nW vs. 0.06 nW). Unsurprisingly, the max power generated decreases dramatically with both thermal and electrical contact resistances, pointing to the urgent need to find strategies to decrease contact resistance for real applications. However, contact resistances cannot be fully eliminated in real devices and miniaturization makes the minimization of contact resistance specially challenging.

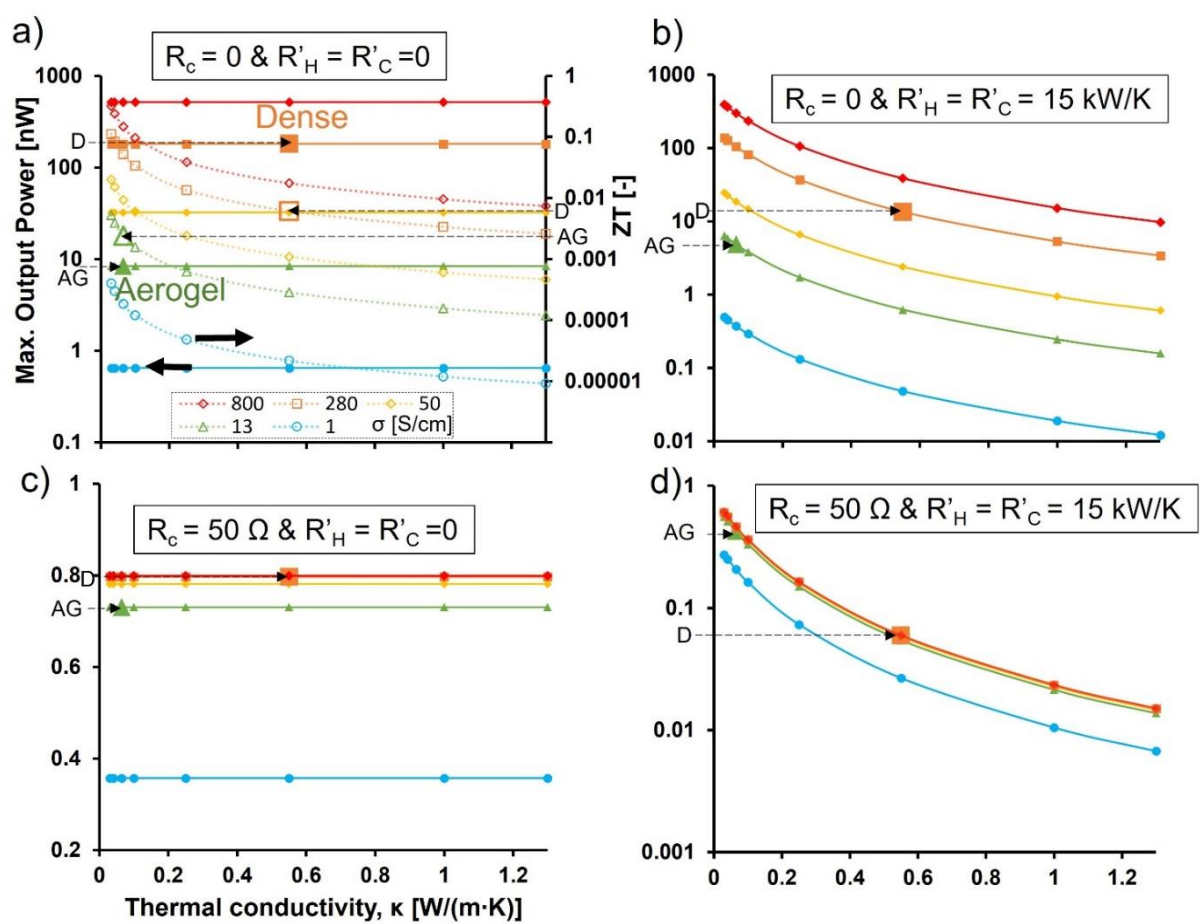

**Figure S10. The maximum output power (left y-axis) and the figure of merit ($ZT$, right y-axis) change for four different cases of electrical and thermal contact resistance.** a) The ideal case: electrical and thermal contact resistances are neglected ($R_C$ = 0 & $R'_H$ = $R'_C$ = 0). b) The thermal contact resistance is taken into account but the electrical contact resistance is neglected ($R_C$ = 0 & $R'_H$ = $R'_C$ = 15 kW/K). c) The electrical contact resistance is taken into account but the thermal contact resistance is neglected ($R_C$ = 50 $\Omega$ & $R'_H$ = $R'_C$ = 0). d) The realistic case: both electrical and thermal contact resistances are taken into account ($R_C$ = 50 $\Omega$ & $R'_H$ = $R'_C$ = 15 kW/K). For the figure: S = 20 µV/K, $\Delta T_E$ = 20 K, $L$ = 3.1 mm and $A$ = 0.5 mm².

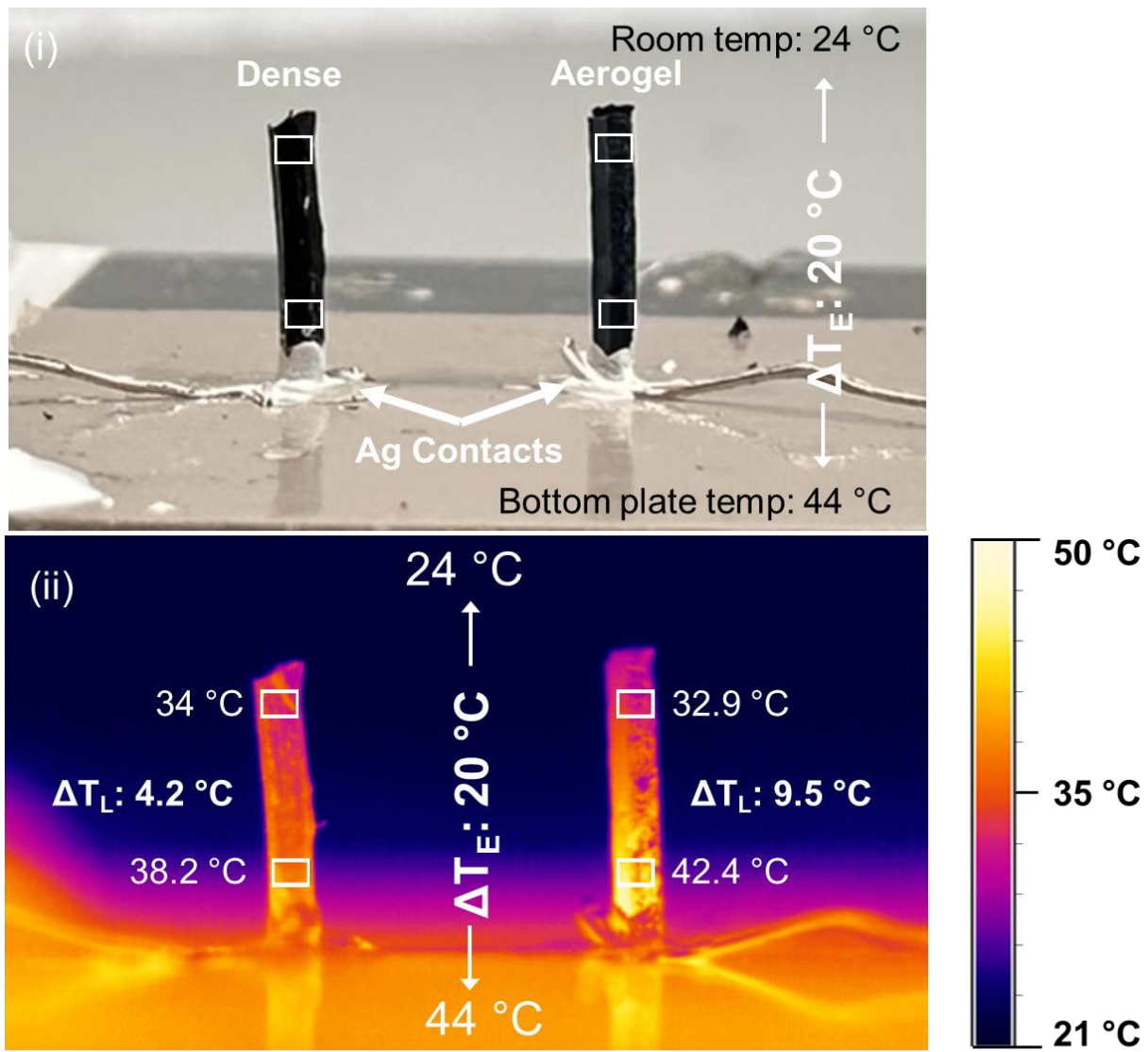

**Figure S11. Comparison of the thermal gradient across a dense and a thermoelectric aerogel pillar.** Optical photograph showing the pillars and indication of the external temperature across the device (ΔT$_E$) (i), and thermal camera image measurement of the actual temperature difference across the TE leg (ΔT$_L$) (ii).

**Table S5. Reported performance of bulky organic thermoelectric materials and through-plane generators.**

| Study | Material | Process | Structure | σ [S/cm] | S [µV/K] | PF [µWm$^{-1}$K$^{-2}$] | κ [W/(m K)] | ZT (x10$^{-3}$) | Device Area [mm$^2$] | Areal Power Density [nW/cm$^2$] |
|---|---|---|---|---|---|---|---|---|---|---|
| **This Work** | **PEDOT:PSS Aerogel** | **3D Printing** | **Small Pillars** | **13.4 ± 0.3** | **22.7 ± 1.4** | **0.69** | **0.065** | **3.2 ± 0.2** | **0.5** | **26** |
| Ref. [10] | PEDOT:PSS Aerogel | 3D Printing | 3D large Objects | 27.15 | 19.34 | 1.02 | 0.075 | 3.99 | 225 | ~0.09 |
| Ref. [11] | PEDOT:PSS Aerogel | 3D Printing | Small Pillars | 71.9 | - | - | - | - | - | - |
| Ref. [12] | PEDOT:PSS Aerogel | Casting | Pellet | 7.9 | 20.3 | 0.33 | 0.04 | 2.4 | - | - |
| Ref. [13] | PEDOT:PSS Aerogel | Casting | Pellet | 2.23 | 17.9 | 0.07 | 0.03 | 0.66 | - | - |
| Ref. [14] | PEDOT:PSS Aerogel | Casting | Pellet | ~0.1 | - | - | - | - | - | - |
| Ref. [15] | PEDOT:PSS Aerogel | Casting | Pellet | 18 ± 1 | 18 ± 1 | 0.59 | 0.065 ± 0.003 | 2.7 ± 0.2 | 132 | ~2.3 |
| Ref. [5] | PEDOT:PSS Aerogel | Casting + dipping | Pellet | ~70 | 18 | ~2.3 | - | - | - | - |
| Ref. [16] | PEDOT:PSS +NMP Compressed Aerogel | Casting and Compressing | Aerogel film | 35 | 18.8 | 1.24 | 0.14 | 2.7 | - | - |
| Ref. [17] | PEDOT:PSS + NFC + GOPS Aerogel | Casting | Pellet | 0.001 | 37 | 0.14 | - | - | - | - |
| Ref. [18] | F4TCNQ doped - P3HT foam | Casting | Pellet | 0.22 | 68.4 | 0.103 | 0.14 | 0.23 | - | - |
| **This Work** | **PEDOT:PSS Dense** | **3D Printing** | **Small Pillars (shrank)** | **281 ± 23** | **17.5 ± 0.4** | **8.6 ± 0.7** | **0.55*** | **4.6 ± 0.4** | **0.5** | **24** |
| Ref. [19] | PEDOT:PSS Dense | Printing in a Cavity | 3D low aspect ratio pillars | ~1 | ~11 | 0.01 | 0.5 | 0.007 | - | ~1 |
| Ref. [20] | CNT-PS | Printing in a Cavity | 3D low aspect ratio pillars | 2.1 | 57 | ~6 | - | - | - | ~5.5 |
| Ref. [21] | PEDOT-Tos | Spin-coating | Thin film | ~70 | ~220 | 324 | 0.37 | 250 | - | - |
| Ref. [22] | PEDOT:PSS – CNT Yarn | Knitting | Fabric | - | - | - | - | - | 3600 | 0.007 |
| Ref. [23] | PEDOT:PSS Yarn | Knitting | Fabric | 43 ± 10 | 14.3 ± 0.7 | 0.88 ± 0.22 | 0.47 | 0.6 ± 0.1 | 2809 | ~0.017 |

*Owing to the isotropic structure of the bulk pillar, the thermal conductivity value was assumed to be the average of the values reported in the in-plane and through-plane directions for films composed of PEDOT:PSS mixed with DMSO[24].